\documentclass[11pt]{article}

\usepackage{amsmath,amssymb,a4wide}
\usepackage{graphicx}

\begin{document}

\title{Kinematic Diffraction from a Mathematical Viewpoint}

\author{{\sc Michael Baake$^{1}$ and Uwe Grimm$^{2}$}\\[2ex]
\normalsize $^{1}$Fakult\"{a}t f\"{u}r Mathematik, 
Universit\"{a}t Bielefeld,\\
\normalsize Postfach 100131, 33501 Bielefeld,
Germany\\[1ex]
\normalsize $^{2}$Department of Mathematics and Statistics, 
The Open University,\\
\normalsize Walton Hall, Milton Keynes MK7 6AA, UK}

\date{}

\maketitle

\begin{abstract}Mathematical diffraction theory is concerned with the
  analysis of the diffraction image of a given structure and the
  corresponding inverse problem of structure determination. In recent
  years, the understanding of systems with continuous and mixed
  spectra has improved considerably. Simultaneously, their relevance
  has grown in practice as well.  In this context, the phenomenon of
  homometry shows various unexpected new facets. This is particularly
  so for systems with stochastic components.

  After the introduction to the mathematical tools, we briefly discuss
  pure point spectra, based on the Poisson summation formula for
  lattice Dirac combs. This provides an elegant approach to the
  diffraction formulas of infinite crystals and quasicrystals.  We
  continue by considering classic deterministic examples with singular
  or absolutely continuous diffraction spectra. In particular, we
  recall an isospectral family of structures with continuously varying
  entropy. We close with a summary of more recent results on the
  diffraction of dynamical systems of algebraic or stochastic origin.\bigskip
\end{abstract}

\clearpage

\tableofcontents

\clearpage

\section{Introduction}

Diffraction, originally observed and investigated in optics in the
17th century, developed into the main tool for structure determination
of solids after X-rays were first used by von Laue and by Bragg in the
early 20th century to determine the atomic structure of crystalline
materials.  Nowadays, a hundred years later, X-ray, electron and
neutron diffraction continue to be among the most important techniques
used in structure analysis (see \cite{C} and references therein),
complemented by direct imaging techniques such as electron and atomic
force microscopy.

The diffraction of a beam of X-rays, electrons or neutrons from a
macroscopic piece of solid is a complicated physical process. The
inverse problem of determining the structure from the diffraction
intensities is even more involved, and rarely determines
the structure uniquely. The presence of inelastic and multiple
scattering, which is particularly prevalent in electron diffraction,
makes it essentially impossible to arrive at a unified mathematical
description of the process.  In contrast, within the realm of
kinematic diffraction in the Fraunhofer limit, there exist powerful
mathematical tools to tackle the direct problem and to gain insight
into the complexity of the corresponding inverse problem.

While the theoretical description had been well developed for the case
of ordinary (periodic) crystals and incommensurate phases, the
discovery of quasicrystals in the early 1980s
\cite{SBGC84,INF85,KN,LevSt} not only required new mathematical methods,
but also re-opened general questions concerning the possible
manifestations of order and disorder in solids. In addition, improved
experimental techniques make the diffuse part of the diffraction
increasingly accessible, and a better understanding of the continuous
diffraction is desirable, in particular in view of the implications on
disorder.
 
In this review, which is an extended version of lectures given by the
authors at RIMS (Kyoto) in summer 2010 \cite{RIMS}, we summarise the
mathematical development that was stimulated by the discovery of
quasicrystals and has led to a more systematic approach to systems 
with all types of spectral components.

\section{Mathematical diffraction}

Since the pioneering paper by Hof \cite{Hof}, it has become evident
that the formulation of mathematical diffraction theory via measures
is both powerful and versatile. As this is not the usual language
of crystallography, we begin with a motivation and then summarise
the mathematical background for this review.

\subsection{Why measures?}

A mathematically satisfying formulation has to talk about infinite
systems. Traditionally, they are often described by functions
(representing the density of the scattering medium) or by tilings
(whose decorations are meant to mimic the atomic positions). These two
points of view are somewhat contradictory in the sense that each
idealisation comes with its own (metric) topology (by which we mean a
notion of `closeness' of structures) which is not compatible with the
other. 

To expand on this, almost periodic functions are compared via the
supremum norm, and one function is close to another when each lives
within an $\varepsilon$-tube around the other, for some small
$\varepsilon>0$. In particular, they are locally $\varepsilon$-close
everywhere. This distance concept makes no sense for the comparison of
two Penrose tilings, say. There, one employs the local topology
instead, which says that two tilings are close when, possibly after a
translation of length at most $\varepsilon$, they agree exactly on a
ball of radius $1/\epsilon$ (while no condition is imposed in the
complement of the ball). This, in turn, makes no sense for almost
periodic functions, because they almost never agree exactly on finite
regions.

This dichotomy has led to rather fierce scientific disputes between
the tiling school and the density function school, in particular in
the years following the discovery of quasicrystals. Mathematically,
however, this is a fight about nothing, because the two viewpoints can
be reconciled by embedding them into a larger, and slightly more
general frame. One possibility to do so is to introduce
\emph{measures} together with the so-called vague topology. Such
measures comprise both almost periodic functions and tilings as
special cases. Moreover, when restricting to one of these cases, the
vague topology agrees with the other topologies mentioned before.

We take this as sufficient motivation and justification to use a
formulation with measures from the very beginning. In addition,
measures provide the most natural frame for systems with disorder, as
is well-known from probability theory. Meanwhile, the relevance of
different topologies has been recognised in a more general setting;
compare \cite{Crelle,M2,BLM} for more. Let us finally mention that a
measure is a mathematical notion that is well suited to quantify both
the distribution of matter and the distribution of (scattered)
intensity in space, so that its appearance in our present
(physically motivated) context is very natural.

\subsection{Mathematical preliminaries}

For simplicity, we introduce measures as linear functionals on
continuous functions, and then connect them to the standard approach
of regular Borel measures via the Riesz-Markov representation theorem;
see \cite{RS} and references therein for background material.

Let $\mathcal{K}=C_{\mathsf{c}}(\mathbb{R}^{d})$ be the space of
complex-valued, continuous functions of compact support. A (complex)
\emph{measure} $\mu$ on $\mathbb{R}^{d}$ is a linear functional on
$\mathcal{K}$ with values in $\mathbb{R}$ (or in $\mathbb{C}$), with
the extra condition that, for every compact set
$K\subset\mathbb{R}^{d}$, there is a constant $a^{}_{K}$ such that
\[
    \lvert \mu(g)\rvert \, \le \, a^{}_{K}\, \lVert g\rVert^{}_{\infty}
\]
for all test functions $g\in\mathcal{K}$ with support in $K$. Here,
$\lVert g\rVert^{}_{\infty}=\sup_{x\in K}\lvert g(x)\rvert$ is the
supremum norm of $g$. If $\mu$ is a measure, the \emph{conjugate} of
$\mu$ is defined by the mapping $g\mapsto\overline{\mu(\bar{g})}$. It
is again a measure, denoted by $\bar{\mu}$. A measure is called
\emph{real} (or signed), when $\bar{\mu}=\mu$, and it is called
\emph{positive} when $\mu(g)\ge 0$ for all $g\ge 0$. For every measure
$\mu$, there is a smallest positive measure, denoted by $\lvert
\mu\rvert$, such that $\lvert\mu(g)\rvert\le\lvert\mu\rvert(g)$ for
all non-negative $g\in\mathcal{K}$. This is called the \emph{total
  variation} (or absolute value) of $\mu$.

A measure $\mu$ is \emph{bounded} (or finite), if
$\lvert\mu\rvert(1)=\lvert\mu\rvert(\mathbb{R}^{d})$ (see below for
the meaning of this notation) is finite. Otherwise, it is called
unbounded, an example of which is given by Lebesgue measure $\lambda$.
The vector space of measures on $\mathbb{R}^{d}$ is given
the \emph{vague topology}, which means that a sequence of measures
$(\mu_{n})^{}_{n\in\mathbb{N}}$ converges vaguely to $\mu$ if
$\lim_{n\to\infty} \mu_{n}(g) = \mu(g)$ in $\mathbb{C}$ for all
$g\in\mathcal{K}$; compare \cite[p.~114]{RS} for more on this topology
in the context of linear functionals. Note that the term `vague' has a
very precise meaning this way, and vague convergence is a natural and
powerful concept of analysis. The measures defined this way are in
one-to-one correspondence with the regular Borel measures on
$\mathbb{R}^{d}$, by means of the Riesz-Markov representation theorem
\cite{RS}. The $\sigma$-algebra of measurable sets is formed by the
Borel sets, which is the smallest $\sigma$-algebra that contains all
open (and hence also all closed) subsets of $\mathbb{R}^{d}$ in its
standard topology. In view of this identification, we write $\mu(A)$
(measure of a set $A\subset\mathbb{R}^{d}$) and $\mu(g)$ (measure of a
function), and use the meaning interchangeably (for instance for
characteristic functions of a set, where $\mu(1^{}_{A})=\mu(A)$).

The distribution of matter in Euclidean $d$-space is described by a
measure $\omega$ on $\mathbb{R}^{d}$, where we assume an infinite
system that is homogeneous and in equilibrium. In most cases, $\omega$
will be \emph{translation bounded}, which means that, for any compact set
$K\subset\mathbb{R}^{d}$, we have
\[
   \sup_{t\in\mathbb{R}^{d}} \,\lvert \omega\rvert (t+K) \; 
    < \; \infty\, .
\]
Moreover, we assume an amenability property of $\omega$, namely the
existence of its \emph{autocorrelation measure} 
\[
   \gamma \, =  \, \gamma^{}_{\omega} 
                  = \, \omega \circledast \widetilde{\omega} 
                 \, := \lim_{R\to\infty}
                \frac{\;\omega|^{}_{R} * \widetilde{\omega|^{}_{R}}\;}
                   {\mathrm{vol} (B_R)} ,
\]
where $B_{R}$ denotes the open ball of radius $R$ around $0\in
\mathbb{R}^{d}$ and $\omega|^{}_{R}$ the restriction of $\omega$ to
$B_{R}$. Given a measure $\mu$, its `flipped-over' version
$\widetilde{\mu}$ is defined via $\widetilde{\mu}(g) =
\overline{\mu(\widetilde{g})}$ for $g\in\mathcal{K}$, where
$\widetilde{g}(x)=\overline{g(-x)}$.  The volume-averaged (or
Eberlein) convolution $\circledast$ is needed because $\omega$ itself
is an unbounded measure, so the direct convolution is not defined. For
instance, if $\lambda$ denotes the standard Lebesgue measure (for
volume), $\lambda\ast\lambda$ is not defined, while
$\lambda\circledast\lambda=\lambda$.  Note that different measures
$\omega$ can share the same autocorrelation $\gamma$. This phenomenon
is called \emph{homometry}, and we shall see explicit examples later
on.

By construction, the measure $\gamma$ is \emph{positive definite} (or
of positive type), which means $\gamma(g * \widetilde{g}\, )\ge 0$ for
all $g\in\mathcal{K}$. As a consequence, $\gamma$ is Fourier
transformable by general results \cite{BF}. The Fourier transform
$\widehat{\gamma}$ exists and is a positive measure, called the
\emph{diffraction measure} of $\omega$. It describes the outcome of
kinematic diffraction by $\omega$ in the sense that $\widehat{\gamma}$
quantifies how much scattering intensity reaches a given volume in
$d$-space. By the Lebesgue decomposition theorem, relative to Lebesgue
measure $\lambda$, there is a unique splitting
\[
   \widehat{\gamma} \; = \; 
            \widehat{\gamma}^{}_{\mathsf{pp} } +
            \widehat{\gamma}^{}_{\mathsf{sc} } +
            \widehat{\gamma}^{}_{\mathsf{ac} }
\]
of $\widehat{\gamma}$ into its pure point part
$\widehat{\gamma}^{}_{\mathsf{pp}}$ (the Bragg peaks, of which there
are at most countably many), its absolutely continuous part
$\widehat{\gamma}^{}_{\mathsf{ac}}$ (the diffuse scattering with
locally integrable density relative to $\lambda$) and its singular
continuous part $\widehat{\gamma}^{}_{\mathsf{sc}}$ (which is whatever
remains). The last contribution, if present, is described by a measure
that gives no weight to single points, but is still concentrated to an
(uncountable!)  set of zero Lebesgue measure. For a proof and
background material, we refer to \cite[Sec.~I.4]{RS}; see also
\cite{Lausanne} and references therein.

\begin{figure}[t]
\centerline{\includegraphics[width=0.9\textwidth]{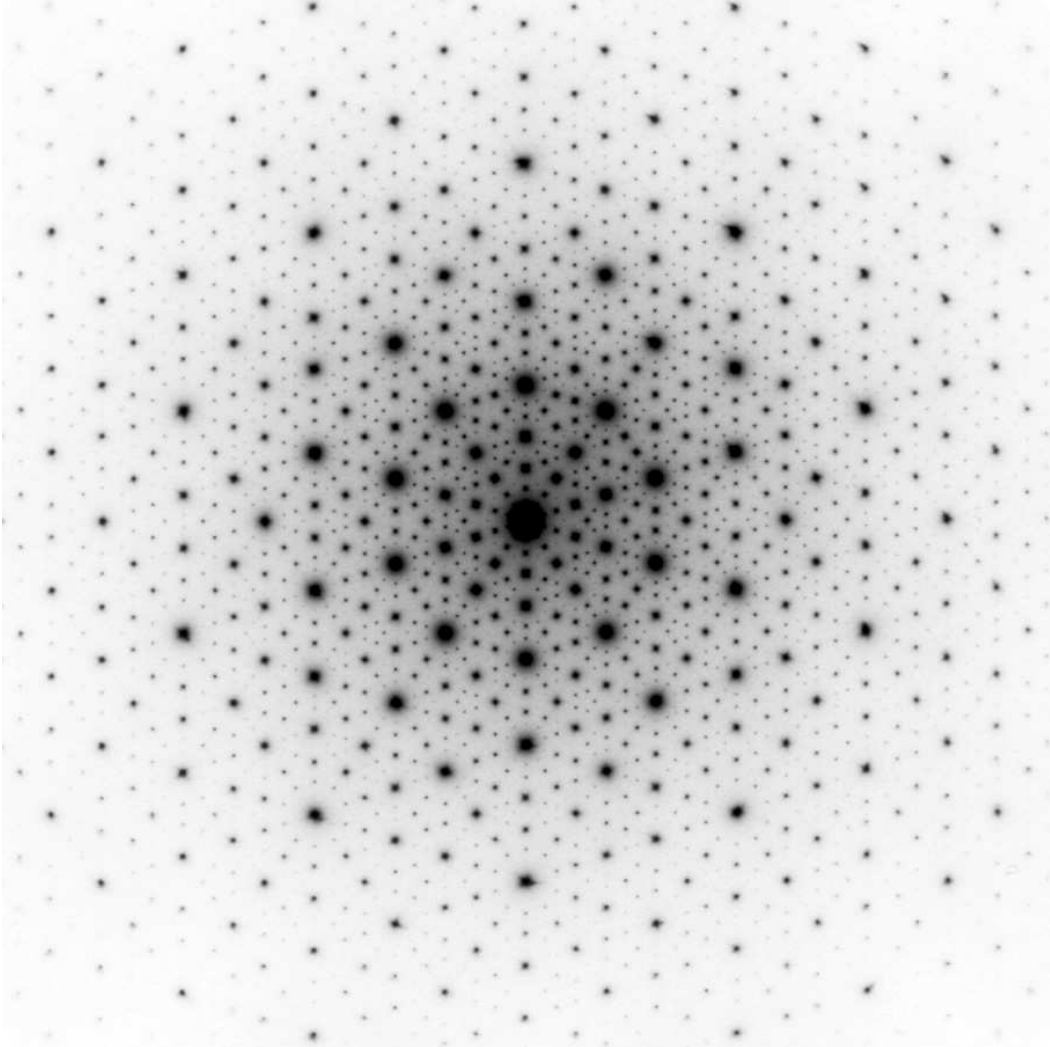}}
\caption{Experimental diffraction pattern of an
  icosahedral \textsf{AlMnPd} quasicrystal along the fivefold
  direction (courtesy C.~Beeli).}
\label{fig:ico}
\end{figure}

Systems with $\widehat{\gamma}= \widehat{\gamma}^{}_{\mathsf{pp} }$
are called \emph{pure point diffractive}. Important examples are
perfect crystals and quasicrystals \cite{L08}, such as the
icosahedrally symmetric \mbox{\textsf{AlMnPd}} alloy that produces the
diffraction image of Figure~\ref{fig:ico}. Mathematical examples of
all spectral types will be discussed below. For general background, we
refer to \cite{BMBook}, and to \cite{BGLit} in particular. The
increasing need for a better understanding of diffuse scattering is
evident from \cite{Withers} and references therein.

It is clear that the diffraction measure is a unique attribute of the
structure described by $\omega$ (under the mild assumption that
$\gamma$ exists, which is a realistic assumption from the physical
point of view, and very natural mathematically). In contrast, the
inverse problem of determining $\omega$ from $\widehat{\gamma}$ is
generally non-unique and a hard problem to solve, both mathematically
and practically.

\subsection{Lattice Dirac combs}

For simplicity, we will, whenever possible, explain the different
scenarios with (weighted) Dirac combs on $\mathbb{Z}$. This means
that, given a bi-infinite sequence $w = (w_{n})^{}_{n\in\mathbb{Z}}$,
we consider
\begin{equation}\label{eq:def-dirac}
  \omega \, = \, w\,\delta^{}_{\mathbb{Z}} \, := \, 
  \sum_{n\in\mathbb{Z}} w(n)\,\delta_{n} ,
\end{equation}
where $\delta_{n}$ is the normalised Dirac measure located at $n$.
(Recall that $\delta_{x}$, for a fixed $x\in\mathbb{R}$, is defined by
$\delta_{x}(g)=g(x)$ for arbitrary $g\in\mathcal{K}$).  The weights
$w(n)=w_{n}$ are assumed to be bounded, and we use the notations
$w(n)$ and $w_{n}$ interchangeably. A simple calculation shows that
the corresponding autocorrelation (which will indeed exist in all examples
discussed here) is then of the form
\begin{equation}\label{eq:autodef}
 \gamma \, = \, \sum_{m\in\mathbb{Z}} \eta(m)\,\delta_{m} , 
\end{equation}
with the autocorrelation coefficients
\begin{equation}\label{eq:eta}
   \eta(m) \, = \, \lim_{N\to\infty} \frac{1}{2N\!+\!1} 
   \sum_{n=-N}^{N} w(n) \, \overline{w(n\!-\!m)}.
\end{equation}
In line with our previous remark, we will assume that the
limit exists for all $m\in\mathbb{Z}$, which is equivalent
to the existence of $\gamma$ in this case. A stochastic setting 
will actually require the existence in a suitable probabilistic
sense; we postpone this point of view until Section~\ref{sec:ac}.

Let us now illustrate the possible spectral types by means of
important and characteristic examples.

\section{Pure point spectra}

The Dirac comb of a general point set $S$ is defined as $\delta^{}_{S}
:= \sum_{x\in S} \delta_{x}$, with $\delta_{x}$ the normalised point
(or Dirac) measure at $x$. If $\varGamma\subset\mathbb{R}^{d}$ is a
\emph{lattice} (meaning a discrete subgroup of $\mathbb{R}^{d}$ such
that the factor group $\mathbb{R}^{d}/\varGamma$ is compact), the
corresponding Dirac comb $\delta_{\!\varGamma}$ itself is Fourier
transformable via the \emph{Poisson summation formula} (PSF)
\begin{equation}\label{eq:psf}
 \widehat{\delta^{}_{\!\varGamma}}
     \, = \, \mathrm{dens} (\varGamma) \, \delta^{}_{\!\varGamma^{*}}\, ,
\end{equation}
where $\varGamma^{*}=\{x\in\mathbb{R}^{d}\mid xy\in\varGamma \text{ for
  all } y\in\varGamma\}$ denotes the dual lattice of $\varGamma$ and
$\mathrm{dens}(\varGamma)$ the density of $\varGamma$; see
\cite{Cordoba,Crelle} and references therein for details. This formula
is fundamental in many ways (and disciplines), and can easily be
proved in two steps; see \cite{Lausanne} for an explicit
exposition. Let us explain what \eqref{eq:psf} means for crystals and
quasicrystals.

\subsection{Crystallographic systems}

A perfect (infinite) crystal with $\varGamma$ as its lattice of
periods can be described by the measure $\omega = \mu *
\delta^{}_{\!\varGamma}$, where $\mu$ is a suitable finite measure. A
convenient (though not unique) choice for $\mu$ is the restriction of
$\omega$ to a fundamental domain of $\varGamma$.  A simple calculation
leads to the autocorrelation
\[
     \gamma = \mathrm{dens} (\varGamma)\, 
      (\mu * \widetilde{\mu}\hspace{1pt})
              * \delta^{}_{\!\varGamma}
\]
because $\widetilde{\delta^{}_{\!\varGamma}}=\delta^{}_{\!\varGamma}$ and
$\delta^{}_{\!\varGamma}\circledast\delta^{}_{\!\varGamma}=
\mathrm{dens} (\varGamma)\,\delta^{}_{\!\varGamma}$. The Fourier transform
of $\gamma$ exists and reads
\[
     \widehat{\gamma} = \bigl(  \mathrm{dens} (\varGamma) \bigr)^{2}
     \, \big| \widehat{\mu} \big|^{2} \, \delta^{}_{\!\varGamma^{*}}
\]
by an application of the convolution theorem together with the PSF
\eqref{eq:psf}. Note that $\big| \widehat{\mu} \big|^{2}$ is a
uniformly continuous and bounded function that is evaluated only at
points of the dual lattice $\varGamma^{*}$. Different admissible
choices for the measure $\mu$ lead to different such functions that
agree on all points of $\varGamma^{*}$, so that the result does not
depend on the choice made. The measure $\widehat{\gamma}$ is a pure
point measure, as one expects for lattice periodic measures $\omega$.

\subsection{Model sets}

Lattice periodic point sets form a special case of the larger class of
regular model sets \cite{M,Martin}, which also lead to pure point
diffraction measures. One of the simplest non-periodic examples in one
dimension emerges from the \emph{silver mean} substitution rule
\[
    \varrho\! : \;  \begin{array}{l} a\mapsto aba\\ b\mapsto a
   \end{array}
\]
which has inflation multiplier $\sigma = 1+\sqrt{2}\,$. The latter is
known as the silver mean and is the Perron-Frobenius eigenvalue of the
corresponding substitution matrix
$\left(\begin{smallmatrix}2&1\\ 1&0\end{smallmatrix}\right)$.  Note
  that $\sigma$ is a Pisot-Vijayaraghavan (PV) number, which means
  that it is an algebraic integer $\sigma>1$ whose algebraic
  conjugates (apart from $\sigma$ itself) all lie inside the open unit
  disk; see \cite{BDGPS} for background material and details. 

\begin{figure}[t]
\centerline{\includegraphics[width=0.9\textwidth]{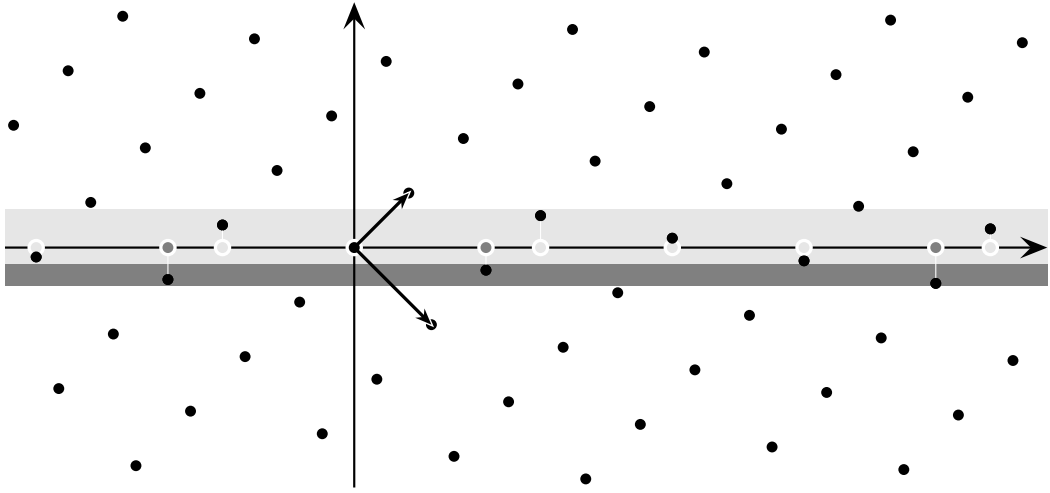}}
\caption{The silver mean point set $\varLambda$
  of Equation~\eqref{eq:silvermean} as a projection of a strip-shaped
  subset of the lattice $\varGamma=\{(x,x^{\prime})\mid
  x\in\mathbb{Z}[\sqrt{2}\,]\}$. The shading marks the left endpoints
  of $a$-type (light grey) and $b$-type (dark grey) intervals.}
\label{fig:silvermean}
\end{figure}

The natural geometric realisation of this system, starting from a
bi-infinite fixed point of $\varrho$ with legal seed $a|a$, is built
via two intervals of length ratio $\sigma$.  If $a$ represents an
interval of length $\sigma$ and $b$ one of length $1$, their left
endpoints constitute the silver mean point set
\begin{equation}\label{eq:silvermean}
   \varLambda = \big\{ x \in \mathbb{Z} [\sqrt{2}\,]
   \,\big|\, x' \in [-\tfrac{\sqrt{2}}{2}, \tfrac{\sqrt{2}}{2}\,]\big\},
\end{equation}
the proof of which is not entirely trivial; it is spelled out in
detail in \cite{TAO}. Here, ${}^{\prime}$ denotes algebraic
conjugation in the quadratic field $\mathbb{Q}(\sqrt{2}\,)$, as
defined by $\sqrt{2}\mapsto-\sqrt{2}\,$. Equation
\eqref{eq:silvermean} has a distinctive geometric meaning which is
illustrated in Figure~\ref{fig:silvermean}.

The algebraic aspects are encoded in a \emph{cut and project scheme}
(CPS), which we state for the general Euclidean case as follows:
\begin{equation}\label{eq:cps}
\renewcommand{\arraystretch}{1.2}\begin{array}{r@{}ccccc@{}l}
   & \mathbb{R}^{d} & \xleftarrow{\,\;\;\pi\;\;\,} 
         & \mathbb{R}^{d} \times \, \mathbb{R}^{m}\!  & 
        \xrightarrow{\;\pi^{}_{\mathrm{int}\;}} & \mathbb{R}^{m} & \\
   & \cup & & \cup & & \cup & \hspace*{-1ex} 
   \raisebox{1pt}{\text{\tiny dense}} \\
   & \pi(\mathcal{L}) & \xleftarrow{\; 1-1 \;} & \mathcal{L} & 
   \xrightarrow{\; \hphantom{1-1} \;} & 
       \pi^{}_{\mathrm{int}}(\mathcal{L}) & \\
   & \| & & & & \| & \\
   &  L & \multicolumn{3}{c}{\xrightarrow{\qquad\quad\quad \;\;
       \,\star\, \;\; \quad\quad\qquad}} 
       &   {L_{}}^{\star} & \\
\end{array}\renewcommand{\arraystretch}{1}
\end{equation}
Here, $\mathcal{L}$ is a lattice in $\mathbb{R}^{d+m}$ with certain
properties that are expressed via the images under the canonical
projections $\pi$ and $\pi^{}_{\mathrm{int}}$. In particular,
$L=\pi(\mathcal{L})$ is a bijective image of $\mathcal{L}$, while
$L^{\star}=\pi^{}_{\mathrm{int}}(\mathcal{L})$ is dense in internal
space $\mathbb{R}^{m}$. Due to these properties, the $\star\,$-map
$x\mapsto x^{\star}$ is well-defined on $L$; see \cite{M} for more. In
the silver mean example, we have a CPS with $d=m=1$,
$L=L^{\star}=\mathbb{Z}[\sqrt{2}\,]$, and the $\star\,$-map is given by
algebraic conjugation, as mentioned above.

In general, a \emph{model set} for a given CPS is a set of the form
\begin{equation}\label{eq:ms}
    \varLambda \, = \,
    \bigl\{  x\in L \mid  x^{\star} \in W \bigr\}
\end{equation}
where $W$ is a relatively compact subset of $\mathbb{R}^{m}$; see
Equation~\eqref{eq:silvermean} for the silver mean case. A model set
$\varLambda$ is \emph{regular} when the boundary $\partial W$ of the
window $W$ has zero Lebesgue measure. The entire setting generalises,
without significant complications, to locally compact Abelian groups
as internal spaces \cite{Meyer,M,Martin}. We will refer to this
freedom later on, where the internal space will be based on the
$2$-adic numbers.

The Dirac combs of regular model sets are pure point diffractive
\cite{Hof,Martin,Crelle}.  This is a substantial theorem for which
three different types of proofs are known. The most common one is
based on the connection to dynamical systems theory
\cite{Hof,Martin,LS}, another on a reformulation via almost periodic
measures; see \cite{Crelle,MS04,S05} and \cite[Sec.~4]{G05}, and
\cite[Lemma~6.6]{Q} for a related result in the context of
subshifts.  An even simpler approach follows a suggestion by Lagarias
and is based on the PSF for the embedding lattice together with Weyl's
lemma on uniform distribution \cite{TAO}. The theorem is also
constructive in the sense that it provides an explicit and computable
formula for the diffraction measure of $\delta^{}_{\!\varLambda}$,
namely
\[
    \widehat{\gamma} = \sum_{k\in L{}_{}^{\circledast}}
         \lvert A(k) \rvert^{2}\, \delta_{k}
\]
with Fourier module $L^{\circledast} = \pi (\mathcal{L}^{*})$ and
amplitudes
\[ 
   A(k) = 
  \frac{\mathrm{dens} (\varLambda)}{\mathrm{vol} (W)}
  \, \widehat{1^{}_{\! W}} (-k^{\star})\, ,
\]
where $1^{}_{\! W}$ is the characteristic function of the window
$W$. This formula has several generalisations
\cite{Martin,Crelle,TAO}, in particular to certain weighted Dirac
combs, which we omit for simplicity.

Model sets are widely used to describe and analyse diffraction images
such as that shown in Figure~\ref{fig:ico}. Although real world
quasicrystals will usually not be pure point diffractive, their
average structure is well captured by this approach. In this regard,
quasicrystals behave pretty much like ordinary crystals.

\begin{figure}
\centerline{\includegraphics[width=0.75\textwidth]{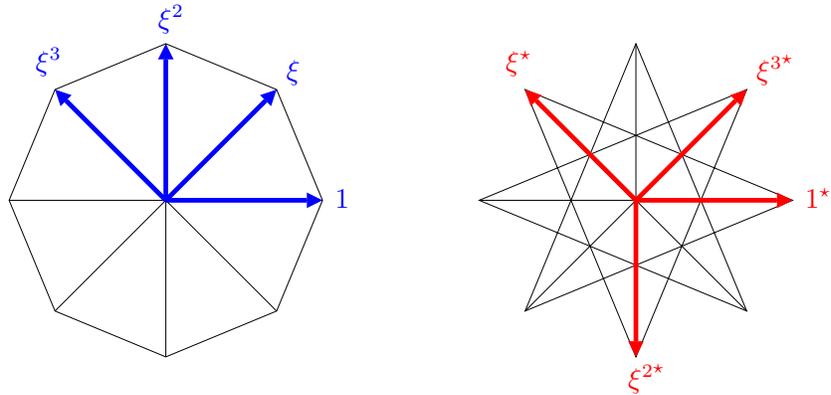}}
\caption{Action of the $\mathbb{Z}$-linear
$\star\,$-map on the generating elements of the eightfold module 
$L = \mathbb{Z} [e^{2\pi \mathrm{i} /8}]$.}
\label{fig:abstars}
\end{figure}

\begin{figure}
\centerline{\includegraphics[height=0.7\textwidth]{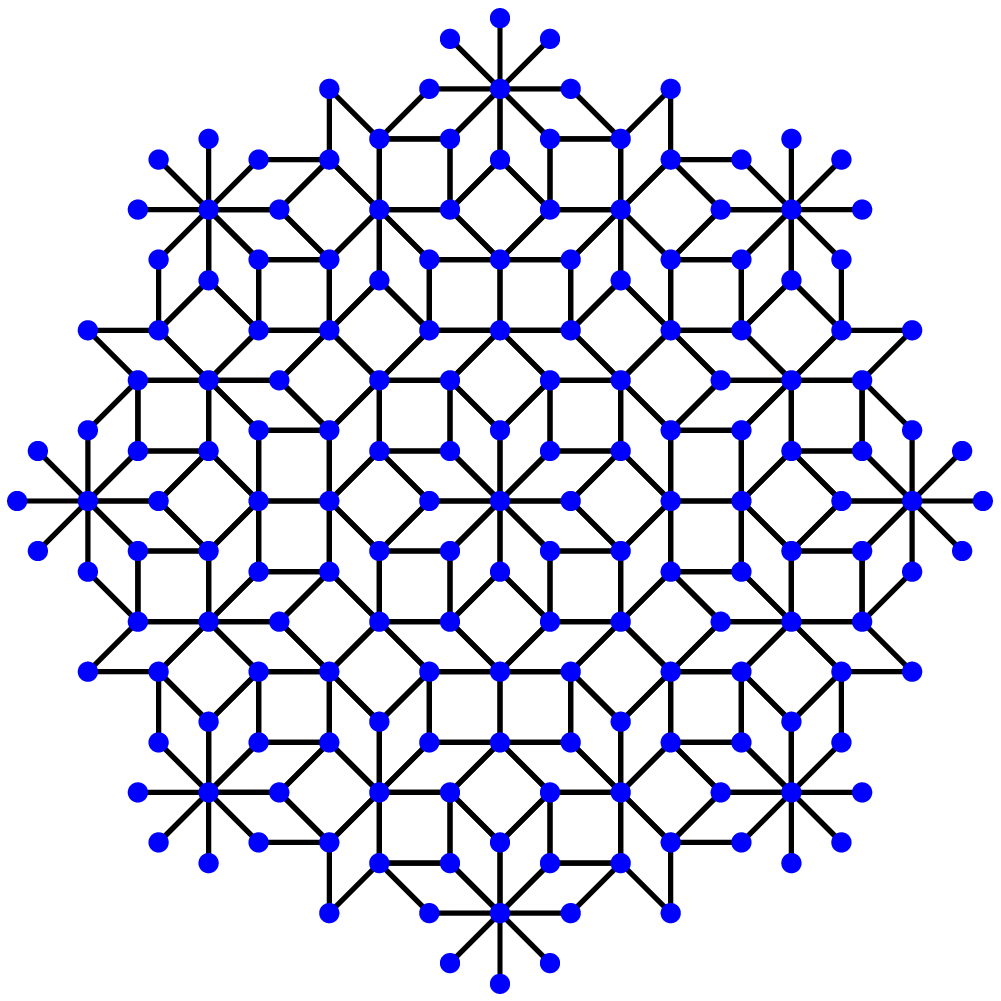}\qquad
\includegraphics[height=0.7\textwidth]{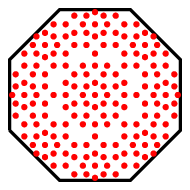}}
\caption{Patch of the Ammann-Beenker tiling and point
  set $\varLambda^{}_{\mathrm{AB}}$ (left) and its lift to internal
  space (right).}
\label{fig:ab}
\end{figure}

Let us expand on the diffraction formula for the vertex set of the
planar Ammann-Beenker (or octagonal) tiling \cite{AGS}. This point set
is a regular model set with $L=\mathbb{Z}[{\xi} ]$, where
$\xi=\exp(2\pi \mathrm{i}/8)$ is a primitive eighth root of unity. The
$\star\,$-map (illustrated in Figure~\ref{fig:abstars}) is defined by
$\xi\mapsto \xi^{3}$, which is an automorphism of the cyclotomic field
$\mathbb{Q}(\xi)$, so that $L^{\star}=L$. The lattice $\mathcal{L} =
\{(x,x^{\star}) \mid x\in\mathbb{Z}[\xi]\}$ is the Minkowski embedding
of $L$, which is a scaled copy of $\mathbb{Z}^{4}$ in this case. The
standard window is a regular octagon $O$ of edge length $1$, centred
at the origin. The model set construction produces the point set
\[
   {\varLambda_{\mathrm{AB}}}   =
   \left\{  x \in \mathbb{Z} 1  +
  \mathbb{Z} \xi  +\mathbb{Z} \xi^2  
    + \mathbb{Z} \xi^3  \mid
   {x^\star} \in{O}\right\}
\]
and its $\star\,$-image $\varLambda_{\mathrm{AB}}^{\star}$ of
Figure~\ref{fig:ab}. The corresponding tiling emerges by connecting
all vertices of distance $1$ in $\varLambda_{\mathrm{AB}}$. It has
two prototiles, a square and a rhombus.

\begin{figure}[t]
\centerline{\includegraphics[width=0.9\textwidth]{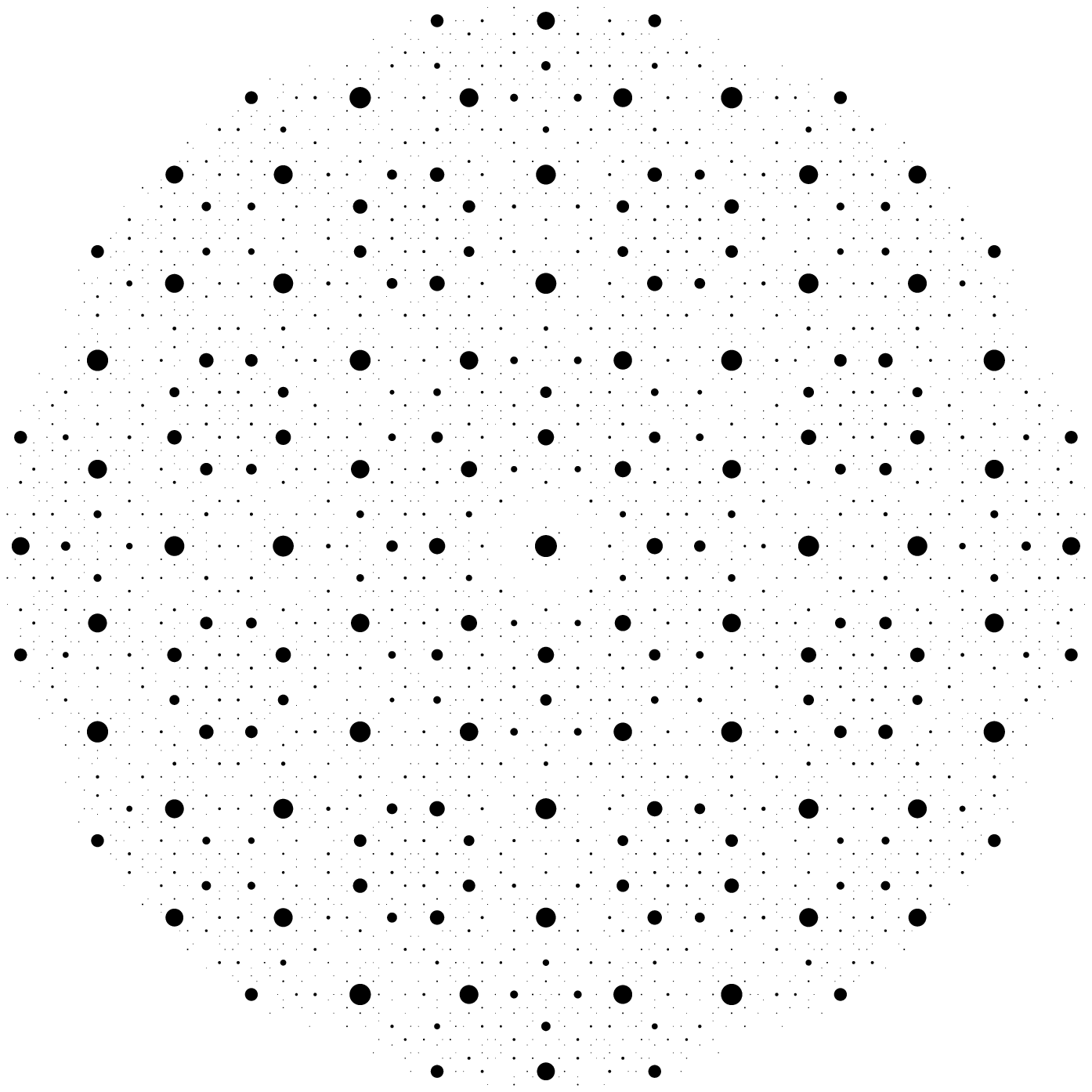}}
\caption{Computed diffraction image of the
  Ammann-Beenker point set $\varLambda^{}_{\mathrm{AB}}$; see 
  text for details.}
\label{fig:abft}
\end{figure}

The diffraction measure is calculated via the Fourier transform of the
characteristic function $1^{}_{O}$. This leads to a dense (but
countable) set of Bragg peaks whose intensities are locally summable.
This means that, in any compact region, there are only finitely many
peaks above any given positive threshold. A precise explicit
calculation can easily be performed by one of the standard computer
algebra packages.  Figure~\ref{fig:abft} shows the result for a
central patch of the Ammann-Beenker diffraction, with cutoff at
$1/1000$ of the central intensity. Here, a Bragg peak is represented
by a small disk whose area is proportional to the intensity and whose
centre is the position of the Bragg peak. The locations of the peaks
above the threshold is a Delone set (in fact, it is even a Meyer
set); see \cite{Nicu10} for a general result in this direction.

A related and very interesting question concerns the spectral type of
systems that are defined by substitutions; see \cite{Q} for
background. More recently, practically useful tests for pure
pointedness of substitution systems have been developed, compare
\cite{AL} and references given there, but the famous Pisot
substitution conjecture (which states that the dynamical system
associated to an irreducible Pisot substitution has pure point
spectrum) and its higher-dimensional generalisations still remain a
mystery. We will not discuss this point of view in what follows, even
though many of our examples will be defined by substitution rules.

Let us close this paragraph with a brief general comment.  A
translation bounded measure $\omega$ also defines a dynamical system
under the translation action of $\mathbb{R}^{d}$; see
\cite{Martin,BL,BLM} for background. If $\omega$ is pure point
diffractive, the corresponding dynamical spectrum is pure point as
well (the converse also being true). This equivalence is well
understood by now \cite{D93,Martin,LMS,BL,G05,LS} (a related result was
already obtained in \cite[Prop.~4.14]{Q}), but does not extend to
general systems with continuous spectral components \cite{ME,BE}, as
we shall see later on.

\subsection{Homometry}

As mentioned previously, different measures may possess the same
autocorrelation and hence the same diffraction. This phenomenon is
called \emph{homometry} \cite{Pat}. Clearly, $\delta_{t}*\omega$ (with
$t\in\mathbb{R}^{d}$) as well as $\widetilde{\omega}$ have the same
diffraction as $\omega$, but non-uniqueness is generally not exhausted
by this. Let us illustrate how it appears already among pure point
diffractive systems.

The simplest situation emerges for periodic Dirac combs on
$\mathbb{Z}$ with rational weights. As an example, 
Gr\"{u}nbaum and Moore \cite{GM95}
constructed homometric Dirac combs of the form
\begin{equation}\label{eq:gm}  
   \omega \, =\, \delta^{}_{6\mathbb{Z}} * 
   \displaystyle\sum_{j=0}^{5} c_{j}\,\delta_{j}
\end{equation}
with positive integer weights $c_{j}$, which are thus
$6$-periodic. The two choices of Table~\ref{tab:gm} lead to the same
autocorrelation. Even worse, these two cases have the same correlation
functions up to $5$th order, and differ only in higher
orders. Nevertheless, the two Dirac combs are substantially
different. Note that the diffraction measure, which is supported on
$\mathbb{Z}/6$, shows systematic extinction (the intensity vanishes on
all points of the form $k=\ell/6$ with $\ell\equiv 2$, $3$ or $4\bmod
6$). Such extinctions are an indication for non-trivial homometry of
pure point diffractive systems.

\begin{table}[h]
  \caption{\label{tab:gm}Integer weights $c_j$ for the two homometric
    Dirac combs built from Equation~\eqref{eq:gm}; see \cite{GM95} for 
    details.}
\centerline{\begin{tabular}{|c|cccccc|}
\hline
$j$     & 0 & 1 & 2 & 3 & 4 & 5 \\
\hline
$c_{j}$ & 11 & 25 & 42 & 45 & 31 & 14 \\
$c_{j}$ & 10 & 21 & 39 & 46 & 35 & 17 \\
\hline
\end{tabular}\bigskip}
\end{table}

\begin{figure}
\centerline{\raisebox{4ex}{
\includegraphics[width=0.44\textwidth]{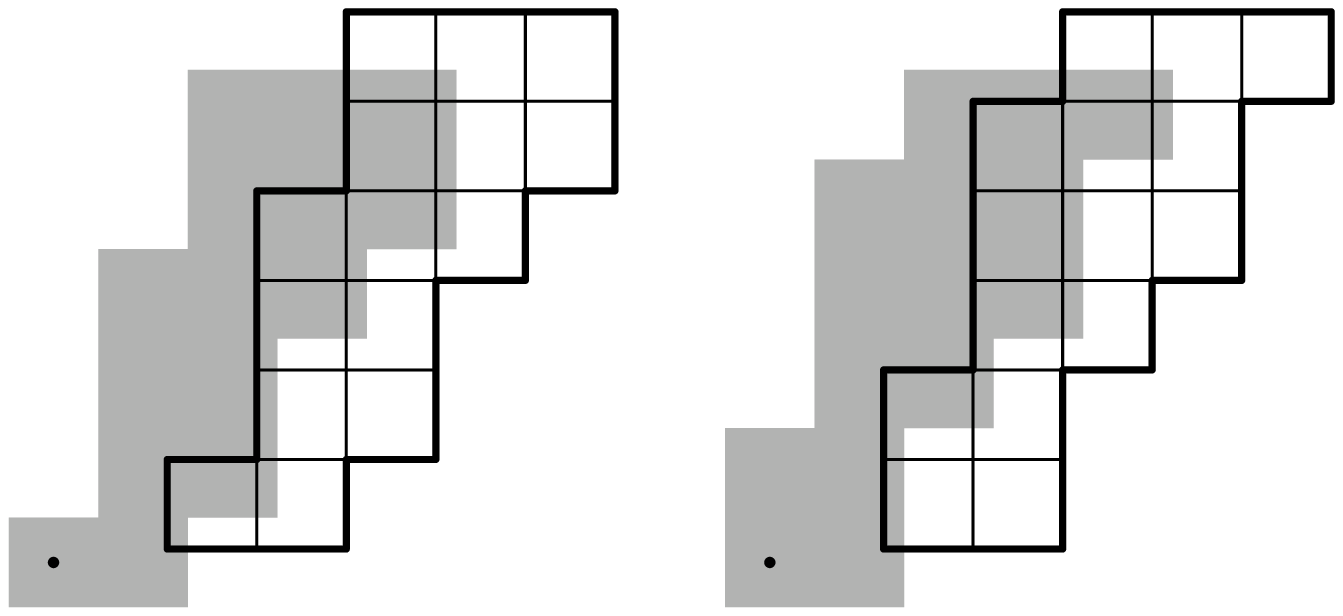}}\qquad
\includegraphics[width=0.37\textwidth]{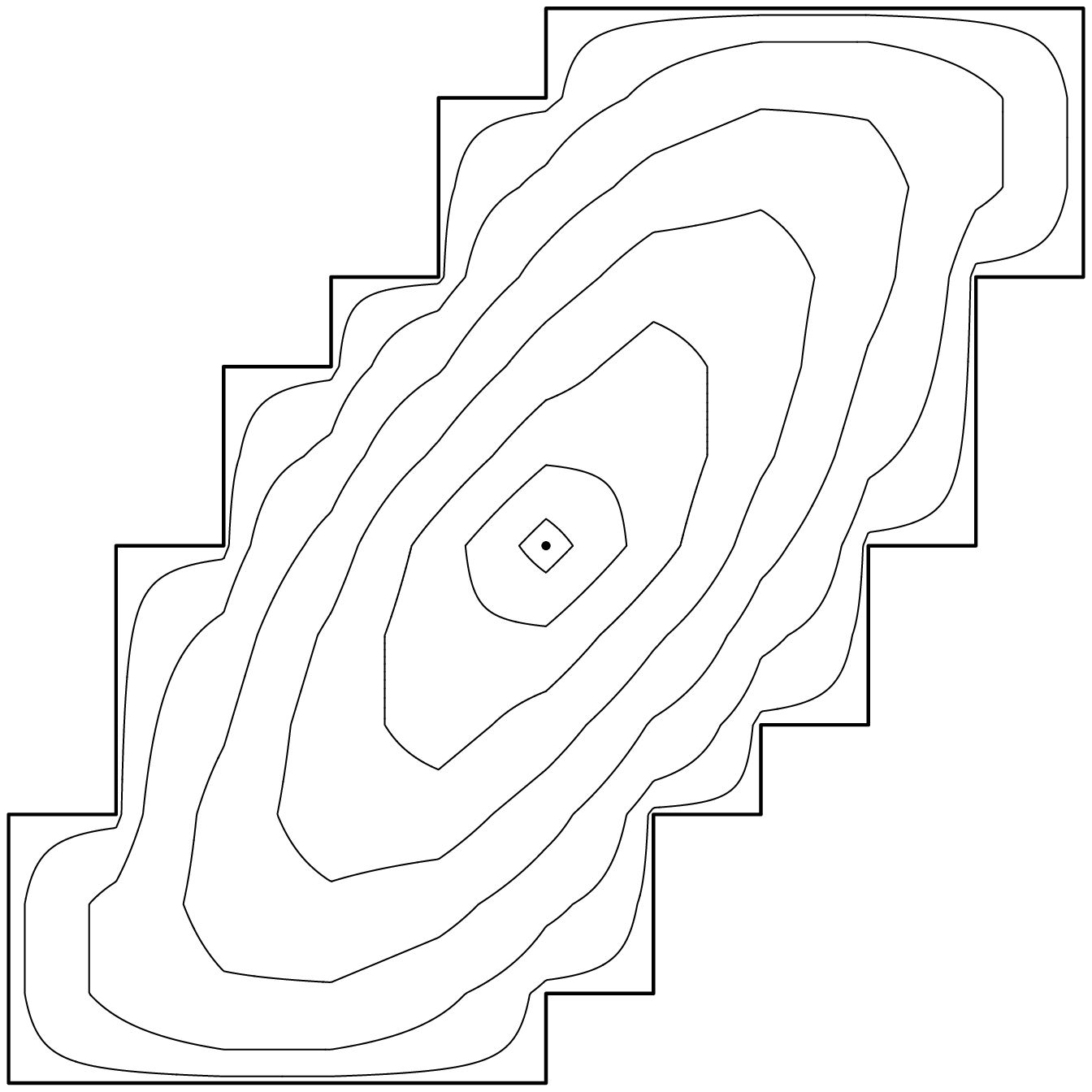}}
\caption{Two polyominoes (left) and their common 
covariogram (right). The small dots mark the origin, see text 
for further details.}
\label{fig:covar}
\end{figure}

Let us go one step beyond and show a pair of homometric model
sets. We use the CPS of the Ammann-Beenker tiling from above, but
replace the octagonal window by one of the two polyominoes shown in
Figure~\ref{fig:covar}, where we follow \cite{BG07}. The Dirac comb
$\delta^{}_{\!\varLambda}$ of the corresponding model set $\varLambda$
with window $W$ has an autocorrelation of the form
\begin{equation}\label{eq:homauto}
    \gamma^{}_{\!\varLambda} \, = \, \sum_{z\in\varLambda-\varLambda}
    \eta(z)\, \delta_{z}\, ,
\end{equation}
where $\varLambda-\varLambda$ is locally finite because $\varLambda$
is a model set. The autocorrelation coefficients in \eqref{eq:homauto}
are given by
\begin{equation}\label{eq:hometa}
     \eta(z) \, = \, \mathrm{dens}(\varLambda)\,
     \frac{\mathrm{vol}\bigl(W\cap (W-z^{\star})\bigr)}{\mathrm{vol}(W)}
     \, = \, \mathrm{dens}(\mathcal{L})\, \mathrm{cvg}^{}_{W}(z^{\star})\, .
\end{equation}
Here, $\mathrm{cvg}^{}_{W}$ is the \emph{covariogram} of $W$, defined by
\[
   \mathrm{cvg}^{}_{W}(x) \, =\, 
    \mathrm{vol}\bigl(W\cap (x+W)\bigr) \, = \,
    \bigl(1^{}_{W} * 1^{}_{-W}\bigr)(x)\, .  
\]
This function is symmetric under reflection in the origin, and one
also has the relations $ \mathrm{cvg}^{}_{t+W} = \mathrm{cvg}^{}_{W}$
for arbitrary $t\in\mathbb{R}^{d}$ as well as $ \mathrm{cvg}^{}_{(-W)}
= \mathrm{cvg}^{}_{W} $. The covariogram for our two polyominoes is
illustrated as a contour plot in Figure~\ref{fig:covar}. The
polyominoes in the same figure are shown with a shifted overlay
structure, which can be used by the reader to check the claimed
homometry (for the displayed shift) on the basis of
Equations~\eqref{eq:homauto} and \eqref{eq:hometa}.

The two model sets constructed this way differ on positions of
positive density. Depending on the length scale of the windows, they
may or may not be locally equivalent via a mutual local derivation
(MLD) rule \cite{B}, but they are always homometric. Further details
are discussed in \cite{BG07,GB08}. If one has access to correlation
functions of higher order, a distinction is possible. In our example,
the $3$-point correlations tell the two examples apart.  Some rather
general results in this direction were recently derived in
\cite{DM09,LM}.

\section{Singular continuous spectra}

\begin{figure}
\centerline{\includegraphics[width=0.8\textwidth]{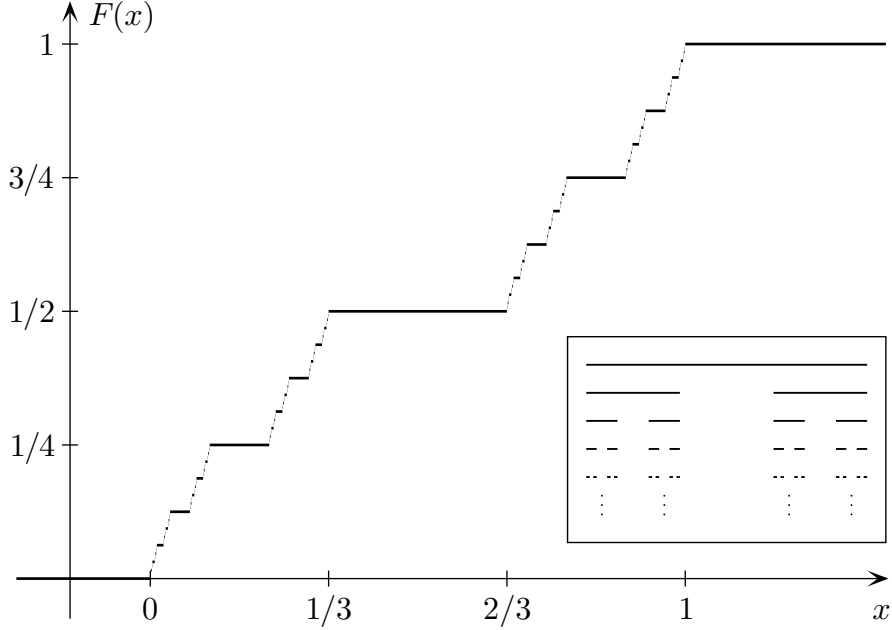}}
\caption{Illustration of the construction of the
  middle-thirds Cantor set $\mathcal{C}$ (inlay) and the distribution
  function $F$ for the corresponding probability measure on $\mathcal{C}$.}
\label{fig:cantor}
\end{figure}

The probably best known singular continuous measure is the one that
emerges from the middle-thirds Cantor set construction. Its
distribution function $F$ is shown in Figure~\ref{fig:cantor}, which
is widely known as the Devil's staircase. This function is continuous
and non-decreasing, but constant almost everywhere. More precisely,
the underlying measure $\mu=\mathrm{d}F$ is concentrated on the Cantor
set $\mathcal{C}$, which is an uncountable set of zero Lebesgue
measure. The (positive) measure $\mu$ is singular continuous, with
$\mu(\{x\})=0$ for all $x\in [0,1]$ and $\mu(\mathcal{C})=1$.

\subsection{Thue-Morse sequence}

Let us now discuss a classic example from the theory of substitution
systems that leads to a singular continuous diffraction measure with
rather different features in comparison with the Cantor measure. This
example has a long history, which can be extracted from
\cite{Wie27,Mah27,K,AS}. We confine ourselves to a brief summary of
the results, and refer to \cite{BG08} and references therein for
proofs and details.

The classic \emph{Thue-Morse} (TM) sequence can be defined via the
one-sided fixed point $v=v^{}_{0}v^{}_{1}v^{}_{2}\ldots$ (with
$v^{}_{0}=1$) of the primitive substitution rule
\[
   \varrho\! : \, 
      \begin{array}{c} 1 \mapsto 1 \bar{1} \\ 
      \bar{1} \mapsto \bar{1} 1 \end{array}
\]
on the binary alphabet $\{1,\bar{1}\}$. The fixed point is the limit
(in the obvious product topology) of the iteration sequence
\[
  1 \stackrel{\varrho}{\longmapsto} 
  1\bar{1}  \stackrel{\varrho}{\longmapsto} 1\bar{1}\bar{1}1
  \stackrel{\varrho}{\longmapsto}  1\bar{1}\bar{1}1 \bar{1}11\bar{1} 
   \stackrel{\varrho}{\longmapsto} \ldots 
  \longrightarrow v = \varrho (v) = v^{}_{0} v^{}_{1} v^{}_{2} 
  v^{}_{3} \ldots 
\]
and has a number of distinctive properties \cite{AS,Q}, for instance
\begin{itemize}
\item       $v^{}_{i} = (-1)^{\text{sum of the binary digits of $i$}}$
\item $v^{}_{2i} \vphantom{\bar{v}} = v^{}_{i}$ and 
        $v^{}_{2i+1} = \bar{v}^{}_{i}$
\item       $v = v^{}_{0} v^{}_{2} v^{}_{4}\ldots$ and
             $\bar{v} = v^{}_{1} v^{}_{3} v^{}_{5}\ldots$
\item       $v$ is (strongly) cube-free.
\end{itemize}
Here, we define $\bar{\bar{1}}=1$ and identify $\bar{1}$ with $-1$ for
the later calculations with Dirac combs.  A two-sided sequence $w$ can
be defined by
\[  
  w_{i} = \begin{cases}
         v_{i} , & \text{for $i\ge 0$,} \\ 
         v_{-i-1}, & \text{for $i<0$,} \end{cases}
\]
which is a fixed point of $\varrho^{2}$, because the seed
$w^{}_{-1}|w^{}_{0} = 1|1$ is a legal word (it occurs in
$\varrho^{3}(1)$) and $w=\varrho^{2}(w)$. The (discrete) hull
$\mathbb{X}=\mathbb{X}^{}_{\mathrm{TM}}$ of the TM substitution is the
closure of the orbit of $w$ under the shift action, which is
compact. The orbit of any of its members is dense in $\mathbb{X}$.  We
thus have a topological dynamical system $(\mathbb{X}, \mathbb{Z})$
that is minimal. When equipped with the standard Borel
$\sigma$-algebra, the system admits a unique shift-invariant
probability measure, so that the corresponding measure theoretic
dynamical system is strictly ergodic \cite{K,Q}.

Any given $w\in\mathbb{X}$ is mapped to a signed Dirac comb (and hence
to a translation bounded measure) $\omega$ via
\[
    \omega\, =\, \sum_{n\in\mathbb{Z}} w_{n}\,\delta_{n}\, .
\] 
We inherit unique ergodicity, and thus obtain an autocorrelation of
the form \eqref{eq:autodef} with coefficients $\eta(m)$ as in
\eqref{eq:eta}. Due to the nature of the fixed point $w$, 
an alternative way to express the coefficients is
\[
     \eta(m)\, = \lim_{N\to\infty}\,
    \frac{1}{N}\, \sum_{n=0}^{N-1} v_n \, v_{n+m}
\]
for $m\ge 0 $ together with $\eta(-m)=\eta(m)$. It is clear that
$\eta(0)=1$, and the scaling relations of $v$ lead to the recursions
\begin{equation}\label{eq:tmrec}
\begin{split}
    \eta(2m) & \, =\,  \eta(m)\qquad \text{and} \\
    \eta(2m\!+\!1) & \, =\,  
    -\frac{1}{2} \bigl( \eta(m) + \eta(m\!+\!1)\bigr),
\end{split}
\end{equation}
which are valid for all $m\in\mathbb{Z}$. In particular, the second
relation, used with $m=0$, implies $\eta(1) = -\tfrac{1}{3}$, which
can also be calculated directly.

Since $\omega$ is supported on $\mathbb{Z}$, the diffraction measure
$\widehat{\gamma}$ is $1$-periodic, which follows from
\cite[Thm.~1]{B02}, but can also be seen from $\gamma=\eta
\delta^{}_{\mathbb{Z}}$. Indeed, this implies $h\gamma=\gamma$ with
the continuous function $h(k)=e^{2\pi i k}$ (which is $1$ on integer
arguments). Taking the Fourier transform and applying the convolution
theorem backwards, one thus obtains $\delta_{1}\ast \widehat{\gamma} =
\widehat{\gamma}$. Moreover, $\widehat{\gamma}$ is of the form
$\widehat{\gamma} = \mu * \delta^{}_{\mathbb{Z}}$, where
\[
     \mu = \widehat{\gamma} \big|_{[0,1)} \quad \text{together with}\quad 
     \eta(m) = \int_{0}^{1} \mathrm{e}^{2 \pi \mathrm{i} m y}\,
     \mathrm{d}\mu (y)\, ,
\]
the latter due to the Herglotz-Bochner theorem
\cite[Thm.~I.7.6]{Katz}. One can now analyse the spectral type of
$\widehat{\gamma}$ via that of the finite measure $\mu$, where we
follow \cite{K}.  Defining $\varSigma(N) = \sum_{m=-N}^{N}
\bigl(\eta(m)\bigr)^2$, a two-step calculation with the recursion
\eqref{eq:tmrec} establishes the inequality $\varSigma (4N) \le
\frac{3}{2} \varSigma (2 N)$ for all $N\in\mathbb{N}$.  This implies
$\lim_{N\to\infty}\varSigma(N)/N=0$, wherefore Wiener's criterion
\cite{Wie27,Katz} tells us that $\mu$ is a continuous measure, so that
$\widehat{\gamma}$ cannot have any pure point component. Note that the
absense of the `trivial' pure point component on $\mathbb{Z}$ is
due to the use of balanced weights, in the sense that $1$ and $-1$ occur
equally frequent, and thus the average weight is zero. 

\begin{figure}[t]
\centerline{\includegraphics[width=0.8\textwidth]{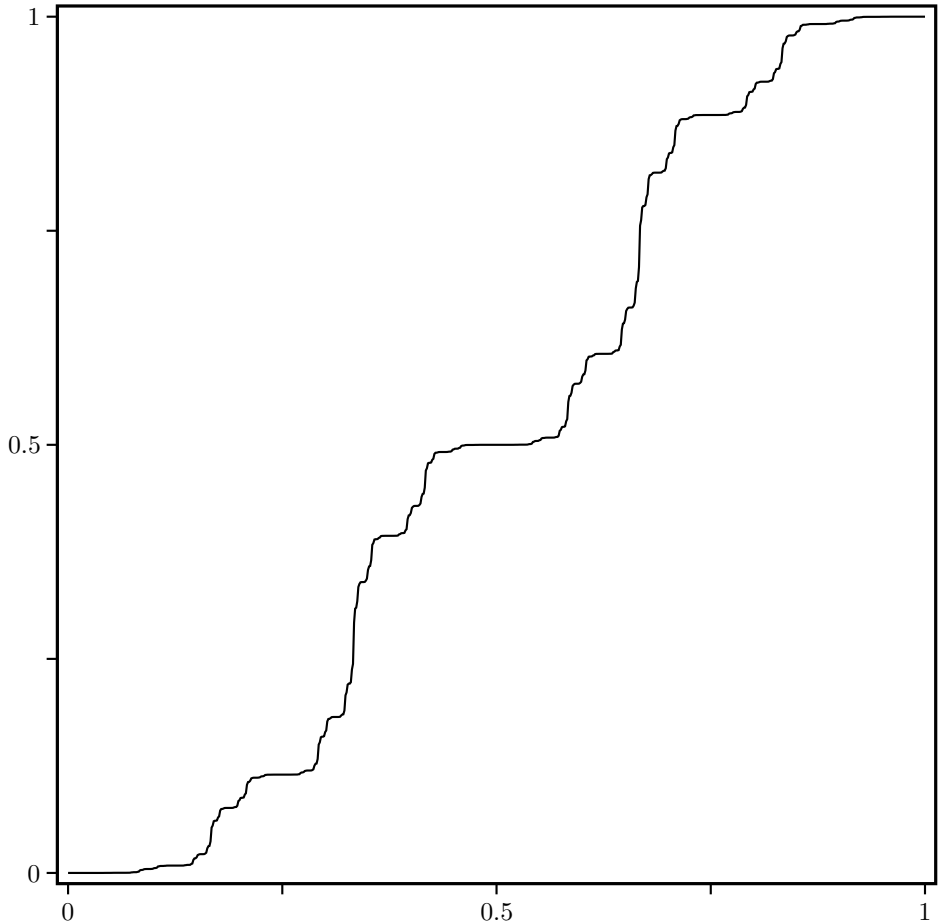}}
\caption{The strictly increasing distribution function
  $F$ of the Thue-Morse diffraction measure on the unit interval.}
\label{fig:tm}
\end{figure}

Let us now define the distribution function $F$ by $F(x) = \mu \bigl(
[0,x] \bigr)$ for $x\in[0,1]$, which is a continuous function that
defines a Riemann-Stieltjes measure \cite[Ch.~X]{Lang}, so that
$\mathrm{d} F=\mu$. The recursion relation for $\eta$ now implies
\cite{K} the functional relations
\[
        \mathrm{d} F \bigl( \tfrac{x}{2} \bigr) \pm
            \mathrm{d} F \bigl( \tfrac{x+1}{2} \bigr) \, = \,
        \left\{\begin{smallmatrix}1 \\
         -\cos(\pi x)\end{smallmatrix}\right\} \,
          \mathrm{d} F (x) \, ,
\]
which have to be satisfied by the $\textsf{ac}$ and $\textsf{sc}$
parts of $F$ separately, because $\mu^{}_{\textsf{ac}} \perp
\mu^{}_{\textsf{sc}}$ in the measure theoretic sense; see
\cite[Thm.~I.20]{RS} or \cite[Thm.~VII.2.4]{Lang}. Therefore, defining
\[  
   \eta_\mathsf{ac} (m)\, =\, \int_{0}^{1} 
    \mathrm{e}^{2 \pi \mathrm{i} m x}\, \mathrm{d}F_{\textsf{ac}}(x)\, ,
\]
we know that the coefficients $\eta^{}_{\mathsf{ac}}(m)$ must satisfy
the same recursions \eqref{eq:tmrec} as $\eta(m)$, possibly with a
different initial condition $\eta^{}_{\mathsf{ac}}(0)$. The
Riemann-Lebesgue lemma \cite[Thm.I.2.8]{Katz}
states $\lim_{m\to\pm\infty} \eta_\mathsf{ac}
(m) = 0$, with is only compatible with $\eta_\mathsf{ac}(0)=0$,
because $\eta_\mathsf{ac}(1)=-\frac{1}{3}\eta_\mathsf{ac}(0)$ and
$\eta_\mathsf{ac}(2m)=\eta_\mathsf{ac}(m)$ for all $m\in\mathbb{N}$,
and hence $\eta_\mathsf{ac}\equiv 0$. This means $F_{\mathsf{ac}}=0$
by the Fourier uniqueness theorem, wherefore $\mu$ and hence
$\widehat{\gamma}$ (neither of which is the zero measure) are purely
singular continuous. The resulting distribution function is
illustrated in Figure~\ref{fig:tm}.  It was calculated by means of the
uniformly converging Volterra iteration
\[
     F^{}_{n+1} (x) = \frac{1}{2} \int_{0}^{2x}
     \bigl( 1 - \cos (\pi y)\bigr) F^{\,\prime}_{n} (y)\,
     \mathrm{d} y  
\]
with $F^{}_{0} (x) = x$. In contrast to the Devil's staircase, the TM
function is \emph{strictly} increasing, which means that there is no
plateau (which would indicate a gap in the support of
$\widehat{\gamma}$); see \cite{BG08} and references therein for
details and further properties of $F$.

Despite the above result, the TM sequence is closely related to
the period doubling sequence, via the (continuous) block map
\begin{equation} \label{eq:block-map}
  \phi \! : \quad 1\bar{1} ,  \bar{1}1 \mapsto a
  \, , \quad 11  ,  \bar{1}\bar{1} \mapsto b \, ,
\end{equation}
which defines an exact 2-to-1 surjection from the hull
$\mathbb{X}^{}_{\mathrm{TM}}$ to $\mathbb{X}^{}_{\mathrm{pd}}$, where
the latter is the hull of the period doubling substitution defined by
\begin{equation}\label{eq:pd-def}
   \varrho^{}_{\mathrm{pd}} \! : \quad
   a \mapsto ab \, , \quad b \mapsto aa\, .
\end{equation}
Viewed as topological dynamical systems, this means that
$(\mathbb{X}^{}_{\mathrm{pd}},\mathbb{Z})$ is a factor of
$(\mathbb{X}^{}_{\mathrm{TM}},\mathbb{Z})$. Since both are strictly
ergodic, this extends to the corresponding measure theoretic dynamical
systems.  The period doubling sequence can be described as a regular
model set with a $2$-adic internal space \cite{BMS,Crelle} and is thus
pure point diffractive. This pairing also explains a phenomenon
observed in \cite{ME}, as the missing part of the \emph{dynamical}
spectrum of the TM system is recovered via the \emph{diffraction}
measure of $\mathbb{X}_{\mathrm{pd}}$.

\subsection{Generalised Morse sequences}

The above example can be generalised to the family 
\[
  \varrho : \, 
      \begin{array}{c} 1 \mapsto 1^{k}\, \bar{1}^{\ell} \\ 
      \bar{1} \mapsto \bar{1}^{k}\, 1^{\ell} \end{array}
\]
with $k,\ell\in\mathbb{N}$, inspired by \cite{Kea68}. They define a
class of systems which we will refer to as the gTM systems.  All
display purely singular continuous diffraction, which follows from
completely analogous arguments \cite{BG10}. The entire analysis is
based on the structure of the autocorrelation, which reads
$\gamma=\eta\,\delta^{}_{\mathbb{Z}}$ with $\eta(0)=1$ and the
recursion relations
\[ 
     \eta \bigl( (k+\ell) m + r\bigr) = \frac{1}{k\!+\!\ell}
     \bigl( \alpha^{}_{k,\ell,r}\, \eta(m) +
     \alpha^{}_{k,\ell,k+\ell-r} \, \eta(m+1) \bigr),
\]
with $\alpha^{}_{k,\ell,r} = k+\ell-r -2 \min (k,\ell,r,k+\ell-r)$.
They are valid for all $m\in\mathbb{Z}$ and $0\le r< k+\ell$. In
particular, one has $\eta\bigl((k+\ell)m\bigr)=\eta(m)$ for
$m\in\mathbb{Z}$. 

\begin{figure}[t]
\centerline{\includegraphics[width=0.48\textwidth]{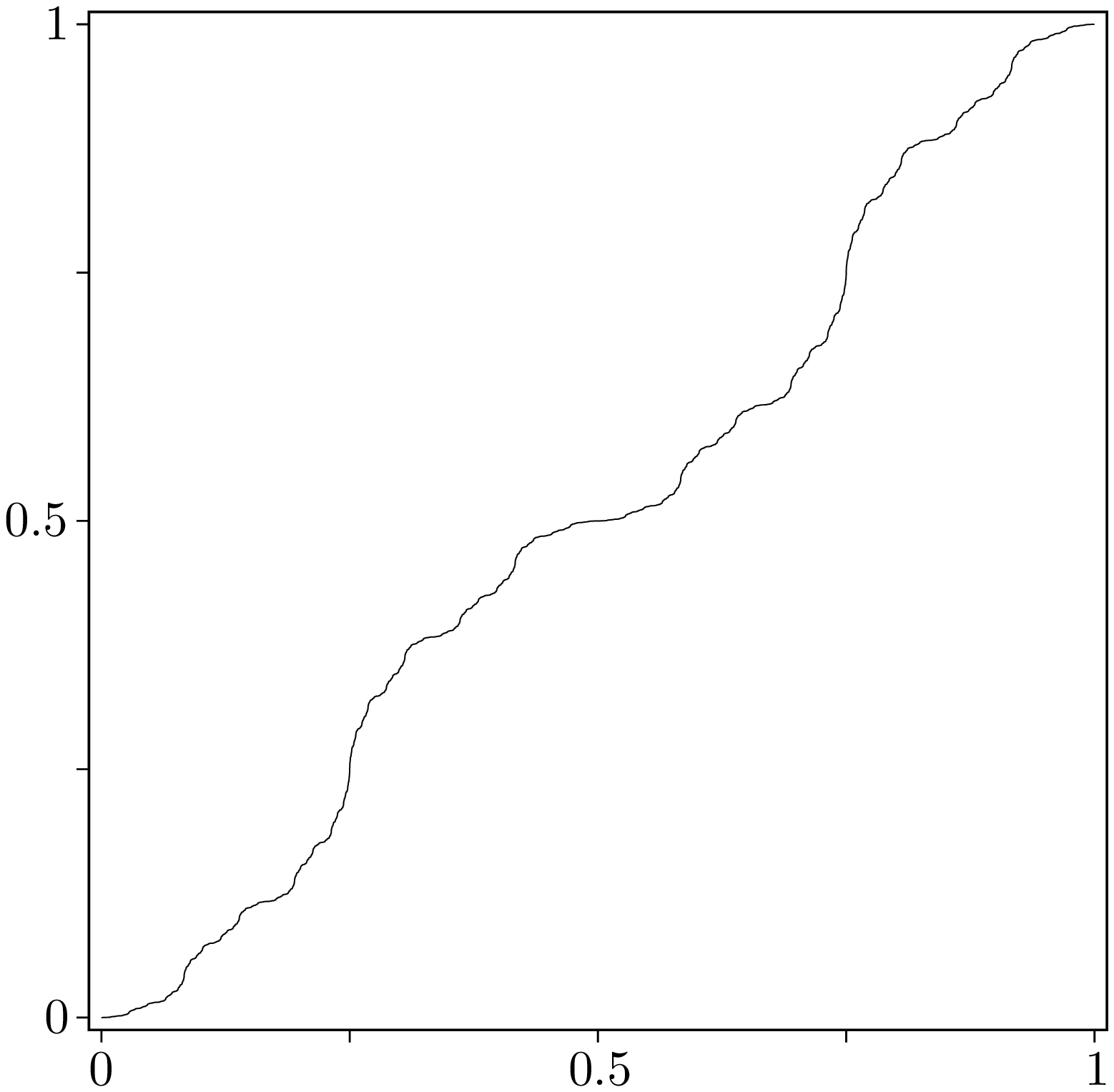} 
   \quad  \includegraphics[width=0.48\textwidth]{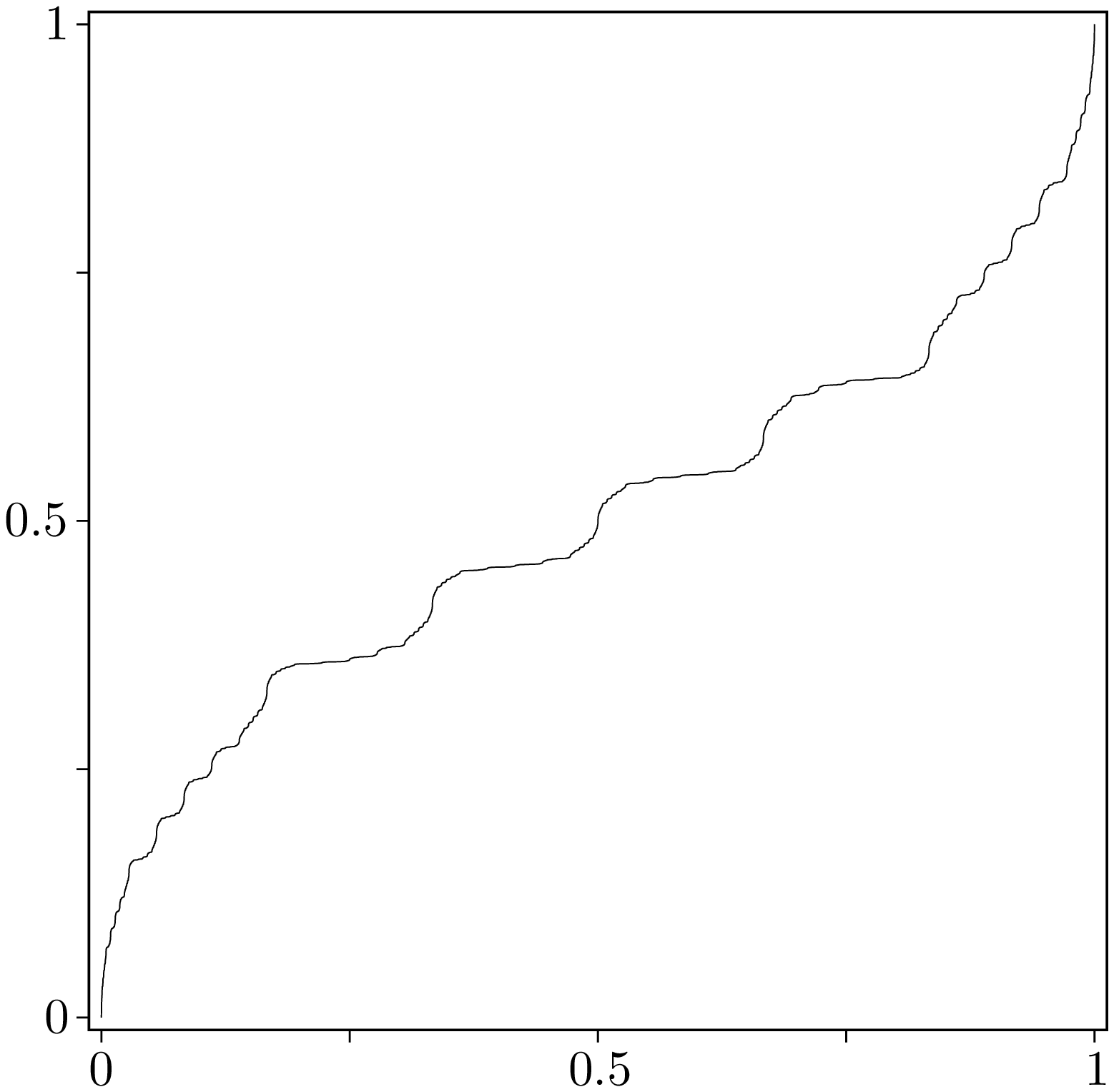}}
\caption{The continuous and strictly increasing
distribution functions of the generalised Morse sequences with
parameters $(2,1)$ (left) and $(5,1)$ (right).}
\label{fig:gtm}
\end{figure}

Given $k,\ell\in\mathbb{N}$, the distribution function $F$ is defined
by $F(x)=\widehat{\gamma}([0,x])$ for $0\le x< 1$, which extends to
$x\in\mathbb{R}$ via $F(x+1)=1+F(x)$. It is also skew-symmetric
($F(-x)=-F(x)$) and thus satisfies $F(q)=q$ for all
$q\in\frac{1}{2}\mathbb{Z}$. The continuous function $F$ possesses the
uniformly converging series expansion \cite{BG10}
\[
       F(x) \,=\,  x + \sum_{m\ge 1} \frac{\eta(m)}{m\pi} \,
    \sin (2\pi m x) \, .
\]
Two further examples are shown in Figure~\ref{fig:gtm}. As mentioned
above, $F$ defines a purely singular continuous measure, for all values
$k,\ell\in\mathbb{N}$.

Another analogy with the TM system is that the block map $\phi$ of
Equation~\eqref{eq:block-map} can still be used to induce a matching
family of generalised period doubling sequences, which turn out
to be conjugate to the Kronecker factor of the gTM system;
compare \cite{Q} for the concept.  These sequences are defined by
the primitive substitution rules
\[ 
  \varrho' : \, 
      \begin{array}{c} a \mapsto b^{k-1}ab^{\ell -1}b \\ 
      b \mapsto  b^{k-1}ab^{\ell -1}a \end{array}
\]
that are all pure point diffractive. The latter claim can most easily
be seen from the letter coincidence (in the sense of Dekking
\cite{Dek78}) at the $k$th position of the images. As before, this
factor explores the pure point part of the dynamical spectrum of the
gTM sequence.

\section{Absolutely continuous spectra}
\label{sec:ac}

The appearance of absolutely continuous diffraction spectra is usually
seen as an indicator for randomness in the structure; see \cite{BLR}
for a discussion in the context of diffraction. Though this is perhaps
generically true, there are also prominent deterministic sequences
with such spectra, such as the Rudin-Shapiro sequence.

In general, within the realm of random structures, one can only expect
almost sure convergence results. In other words, most statements
become measure theoretic in nature, though they are still completely
rigorous. To tame randomness by means of strict laws is one of the
tasks of probability theory, at which it is pretty successful; see
\cite{F1,F2} for general background. Still, this point of view (which
can be rather counterintuitive at times) is something one has to get
used to.

\subsection{Coin tossing sequence}

The simplest example emerges from repeated coin tossing. Here, one
obtains sequences $w\in\{\pm 1\}^{\mathbb{Z}}$ (for instance with $1$
for `head' and $-1$ for `tail') which may be considered as the outcome
of an eternal coin tosser. In more modern (and slightly more general)
terminology, one considers a family
$\bigl(W_{n}\bigr)_{n\in\mathbb{Z}}$ of independent and identically
distributed (i.i.d.) random variables with values in $\{\pm 1\}$ and
probabilities $p$ (for $1$) and $1\!-\!p$ (for $-1$). The ensemble of
all possible realisations is $\mathbb{X}^{}_{\mathrm{B}}=\{\pm
1\}^{\mathbb{Z}}$, which is equipped with a probability measure
$\mu^{}_{\mathrm{B}}$ that emerges from the elementary probabilities
via independence \cite{F1}. This gives a measure theoretic dynamical
system that is called the \emph{Bernoulli shift} \cite{W}. It has
(metric) entropy $H(p) = - p\log (p) - (1\!-\!p) \log (1\!-\!p)$.

A random sequence $W$ leads to a Dirac comb $\omega = W
\delta^{}_{\mathbb{Z}}$, which is now a translation bounded random
measure with support $\mathbb{Z}$. Its autocorrelation, if it exists,
is of the form $\gamma^{}_{\mathrm{B}} =
\eta^{}_{\mathrm{B}}\,\delta^{}_{\mathbb{Z}}$ with
\[
   \eta^{}_{\mathrm{B}}(m) \, := \lim_{N\to\infty} 
   \frac{1}{2N\!\!+\!1}\!\sum_{n=-N}^{N}\! W_{n}W_{n+m} 
    \,\stackrel{\text{\tiny (a.s.)}}{=}\,
   \begin{cases}
   1,  & m = 0, \\
   (2p\! -\! 1)^2, & m\ne 0.
    \end{cases}
\]
Here, the convergence (for $m\ne 0$) is almost sure by the strong law
of large numbers (SLLN) \cite{E}, which means that this is the result
for $\mu^{}_{\mathrm{B}}$-almost all elements of $\{\pm
1\}^{\mathbb{Z}}$; see also \cite{BM98} for further results on the
basis of this type of argument. Note that the use of the SLLN can
also be replaced by an application of Birkhoff's ergodic theorem,
because the Bernoulli shift is ergodic \cite{W}. The corresponding
diffraction measure reads
\[
   \widehat{\gamma^{}_{\mathrm{B}}}\,    \stackrel{\text{\tiny (a.s.)}}{=} 
    \, (2p-1)^{2}\delta^{}_{\mathbb{Z}} \,+\, 
   4 p (1-p)\,\lambda\, ,
\]
which follows from $\gamma^{}_{\mathrm{B}}$ by an application of the
PSF \eqref{eq:psf} together with
$\widehat{\delta^{}_{0}}=\lambda$. For the fair coin
($p=\frac{1}{2}$), this simplifies to
$\widehat{\gamma^{}_{\mathrm{B}}}=\lambda$, which is thus our first
example of a purely absolutely continuous diffraction measure.  Note
that the absence of a pure point component is a result of the (almost
sure) balance between the weights $1$ and $-1$, which is the
probabilistic counterpart of the (deterministic) balance in the TM
example. For a more general discussion of pure point spectra and entropy,
we refer to \cite{BLR} and references therein.

\subsection{Rudin-Shapiro sequence}

\begin{figure}[t]
\centerline{\includegraphics[width=\textwidth]{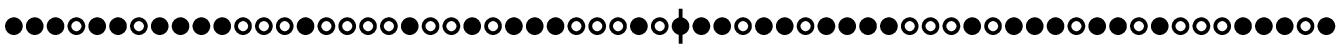}}
\caption{Illustration of the binary Rudin-Shapiro sequence.}
\label{fig:rs}
\end{figure}

Let us contrast the coin tossing sequence with a deterministic
example that derives from \cite{R,S}. This sequence was originally
constructed to show that the absence of pair correlations does not
imply the presence of randomness. This has interesting consequences
in diffraction theory, as pointed out in \cite{HB00}. The modern
formulation of the system is based on the substitution
\[
   \varrho:\quad a\mapsto ac,\quad b\mapsto dc,\quad 
    c\mapsto ab,\quad d\mapsto db\, .
\]
Since $b|a$ is a legal seed (it occurs in $\varrho^{2}(b)$), one can
construct a bi-infinite sequence $u$ by the usual iteration procedure
as
\[
   b|a \; \stackrel{\varrho^{2}}{\longmapsto} \;   
             dbab|acab \; \stackrel{\varrho^{2}}{\longmapsto} \; \ldots \;
            \longrightarrow \; u=\varrho^{2}(u)\, ,
\]
where convergence is in the standard product topology.  The hull
(orbit closure) of $u$ defines the \emph{quaternary} Rudin-Shapiro
system. Its reduction to a binary system is achieved by the mapping
\[
    \varphi:\quad a,c\mapsto 1\, ,\quad b,d\mapsto \bar{1}\, ,
\]
and the orbit closure of $w:=\varphi(u)=\ldots
\bar{1}\bar{1}1\bar{1}|111\bar{1}\ldots$ defines the hull
$\mathbb{X}^{}_{\mathrm{RS}}$ of the \emph{binary} Rudin-Shapiro
system.  The sequence $w$ is illustrated in Figure~\ref{fig:rs}. For
the equivalent description as a weighted Dirac comb on $\mathbb{Z}$, we
again use the identification of $\bar{1}$ with $-1$.

An alternative description of $w$ uses the initial conditions
$w(-1)=-1$, $w(0)=1$ together with the recursion
\begin{equation}\label{eq:rsrec}
   w(4n+\ell)=
    \begin{cases} w(n),  & \mbox{for $\,\ell\in\{0,1\}$,} \\
          (-1)^{n+\ell}\,w(n), & \mbox{for $\,\ell\in\{2,3\}$.}
     \end{cases}
\end{equation}
The autocorrelation of the corresponding weighted Dirac comb
$\omega^{}_{\mathrm{RS}}$ exists and turns out to be
$\gamma^{}_{\mathrm{RS}}=\delta^{}_{0}$. To prove this, one defines
the coefficients
\[
\begin{array}{@{}c@{}} \eta(m)\\ \vartheta(m)\end{array} \bigg\} := 
{\displaystyle\lim_{N\to\infty} \frac{1}{2N\!\!+\!1}\!\sum_{n=-N}^{N}}\! 
w(n)\, w(n+m) \,\bigg\{ \begin{array}{@{}c@{}}1\\ (-1)^{n} . \end{array}
\]
An application of Birkhoff's ergodic theorem to the quaternary
Rudin-Shapiro system (which is strictly ergodic \cite{Q}) establishes
the existence of all these limits.  The recursion \eqref{eq:rsrec}
now implies the recursion relations \cite{BG10}
\[
   \begin{split}
   \eta(4m)& \,=\, \tfrac{1+(-1)^m}{2}\,\eta(m) ,\\[0.5ex]
   \eta(4m\!+\!1) &\,=\, \tfrac{1-(-1)^m}{4}\,\eta(m) +
   \tfrac{(-1)^m}{4}\,\vartheta(m) - \tfrac{1}{4}\,
   \vartheta(m\!+\!1) , \\[0.5ex]
   \eta(4m\!+\!2) &\,=\, 0,\\[0.5ex]
   \eta(4m\!+\!3) &\,=\, \tfrac{1+(-1)^m}{4}\,\eta(m\!+\!1) -
   \tfrac{(-1)^m}{4}\,\vartheta(m)
          + \tfrac{1}{4}\vartheta(m\!+\!1),
   \end{split}
\]
together with
\[
   \begin{split}
   \vartheta(4m) &\,=\, 0,\\[0.5ex]
   \vartheta(4m\!+\!1) &\,=\, \tfrac{1-(-1)^m}{4}\,\eta(m) -
   \tfrac{(-1)^m}{4}\,\vartheta(m) +\tfrac{1}{4}\,\vartheta(m\!+\!1) , \\[0.5ex]
   \vartheta(4m\!+\!2)&\,=\, \tfrac{(-1)^m}{2}\,\vartheta(m)+\tfrac{1}{2}\,
   \vartheta(m\!+\!1) , \\[0.5ex]
   \vartheta(4m\!+\!3) &\,=\, -\tfrac{1+(-1)^m}{4}\,\eta(m\!+\!1) -
   \tfrac{(-1)^m}{4}\,\vartheta(m)
           + \tfrac{1}{4}\,\vartheta(m\!+\!1) .
   \end{split}
\]
with initial conditions $\eta(0)=1$ and $\vartheta(0)=0$. The set of
equations closes in the sense that no new quantities occur. This was
the reason for the introduction of the signed coefficients
$\vartheta(m)$, which show up in the first block of equations. The
system has the unique solution $\vartheta\equiv 0$ together with
$\eta(m)=\delta_{m,0}$, hence $\gamma^{}_{\mathrm{RS}}=\delta^{}_{0}$.

The diffraction measure for the binary Rudin-Shapiro system is thus
given by $\widehat{\gamma}^{}_{\mathrm{RS}} = \lambda$, which
coincides with that of the coin tossing sequence for $p=\frac{1}{2}$;
see also \cite{HB00}.  These two examples are thus homometric, despite
the fact that one is deterministic (with entropy $0$) while the other
is fully stochastic (with entropy $\log (2)$); see \cite{BG09,BG10}
for further details and discussions.

\subsection{Bernoullisation}

The homometry between the Dirac combs of the Rudin-Shapiro and
the balanced coin tossing sequence raises the question how `bad' the
non-uniqueness of the inverse problem in this case really is. A
partial answer (to the negative) can be given by means of the
`Bernoullisation' procedure which was introduced in \cite{GB08,BG09}.

Starting from a uniquely ergodic bi-infinite sequence $S\in\{\pm
1\}^{\mathbb{Z}}$, its Dirac comb $\omega^{}_{S} = S\,
\delta^{}_{\mathbb{Z}}$ possesses a unique autocorrelation
$\gamma^{}_{S}$. Let us now consider the random Dirac comb
\[
   \omega = \sum_{n\in\mathbb{Z}} S_{n}W_{n}\,\delta_{n}\, ,
\]
where $\bigl( W_{n}\bigr)_{n\in\mathbb{Z}}$ is once again an i.i.d.\
family of random variables with values in $\{\pm 1\}$ and
probabilities $p$ and $1-p$. Another (slightly more complicated)
application of the SLLN shows that $\omega$ almost surely has the
autocorrelation
\[
   \gamma    \stackrel{\text{\tiny (a.s.)}}{=} 
   (2p-1)^{2}\,\gamma^{}_{S} + 
   4 p (1-p)\, \delta^{}_{0}\, .
\]
If $S$ is the binary Rudin-Shapiro sequence, which is uniquely ergodic,
a short calculation reveals that 
\[
   \gamma\, =\, \delta^{}_{0}\, = \,\gamma^{}_{\mathrm{RS}}
\]
in this case, \emph{irrespective} of the value of the parameter $p\in
[0,1]$. This way, we find a one-parameter family of homometric (or
isospectral) structures whose entropy varies continuously between $0$
and $\log (2)$.  The conclusion is that kinematic diffraction alone
cannot distinguish order from disorder here \cite{BG09}.

\section{Further directions}

After our brief discourse on the different spectral types of
diffraction measures (by means of some paradigmatic examples), this
section aims at indicating some more recent developments, which 
will again be explained informally by means of selected examples.

\subsection{Ledrappier's model}

Let us consider a prominent planar example of algebraic origin,
which is due to Ledrappier \cite{L}. It is defined as
\[
  \mathbb{X}_{\mathrm{L}} =
  \bigl\{ w \in \{\pm 1\}^{\mathbb{Z}^{2}} \! \mid
     w_{x} w_{x+e^{}_{1}} w_{x+e^{}_{2}} = 1 \,
     \mbox{ {\footnotesize for all} } x \in \mathbb{Z}^{2} \bigr\},
\]
where $e^{}_{1}$ and $e^{}_{2}$ denote the standard Euclidean basis
vectors in the plane. $\mathbb{X}_{\mathrm{L}}$ is a closed subset of
the full shift $\{\pm 1\}^{\mathbb{Z}^{2}}$ and hence compact. It is
also an Abelian group (under pointwise multiplication in our
formulation, which follows \cite{BW10}). As a dynamical system, it is
thus equipped with the corresponding Haar measure
$\mu^{}_{\mathrm{L}}$, which is positive and normalised so that
$\mu^{}_{\mathrm{L}}(\mathbb{X}^{}_{\mathrm{L}})=1$. Obviously, the
system has no entropy, because the knowledge of a configuration along
one horizontal line determines everything above it. However, it is
clearly not deterministic.  In fact, essentially along any given
lattice direction, it looks like a one-dimensional Bernoulli system
\cite{L}. It is thus said to have rank $1$ entropy, which essentially
means that the number of circular patches of a given size grows
exponentially with its diameter, but not with its area.

Given an element $w\in\mathbb{X}_{\mathrm{L}}$, the corresponding
Dirac comb
\[
   \omega \, = \, \sum_{x\in\mathbb{Z}^{2}} w_{x} \, \delta_{x}
\]
possesses $\mu^{}_{\mathrm{L}}$-almost surely the autocorrelation
$\gamma$ and the diffraction measure $\widehat{\gamma}$ given by
\cite{BW10}
\[
  \gamma = \delta^{}_{0}\quad \text{and}\quad
  \widehat{\gamma} = \lambda\, .
\]
The system is thus homometric with the two-dimensional Bernoulli
system with $p=\frac{1}{2}$ (coin tossing on $\mathbb{Z}^{2}$), and
also with the direct product of two binary Rudin-Shapiro
sequences. The similarity with the Bernoulli system goes a lot
further, in the sense that also other correlation functions agree,
although the systems differ for certain $3$-point correlations; see
\cite{BW10} for details and extensions.

This system is meant to indicate that higher-dimensional symbolic
dynamics is good for a surprise, as is well known from \cite{Sch}.  It
is thus clear that the inverse problem becomes more complicated with
growing dimension, as new phenomena show up. Another famous example is
the $(\times 2, \times 3)$ dynamical system, which shares almost all
correlation functions with a Bernoulli system with continuous
($\mathbb{S}^{1}$) degree of freedom \cite{BW10}.

\subsection{Random dimers on the integers}

Back to one dimension, let us briefly describe a system that was
recently suggested by van Enter \cite{BE}. First, consider
$\mathbb{Z}$ as a close-packed arrangement of `dimers' (pairs of
neighbours), hence without gaps or overlaps. There are two
possibilities to do so. Next, give each pair a random orientation by
decorating it with either $(+,-)$ or $(-,+)$, with equal
probability. Identifying $\pm$ with $\pm 1$, this defines the closed
(and hence compact) set
\[
   \mathbb{X} \, = \, \bigl\{ w \in \{ \pm 1 \} ^{\mathbb{Z}} \mid
   M(w) \subset 2 \mathbb{Z} \,\text{ or }
   M(w) \subset 2 \mathbb{Z} + 1 \bigr\} \, ,
\]
where $M(w) := \{ i\in\mathbb{Z} \mid w_{i} = w_{i+1} \}$. Note that
$M(w)$ can be empty, which happens precisely for the two periodic
sequences $\ldots + - | + - \ldots$ and $\ldots - + | - + \ldots$. One
has an invariant measure on $\mathbb{X}$ that emerges from the
stochastic process via the probability of the possible finite patches
(which define the generating cylinder sets as usual).

Turning a configuration $w\in\mathbb{X}$ into a signed Dirac comb with
weights $w_{i}\in\{\pm 1 \}$ as before, another exercise with the SLLN
shows that its autocorrelation almost surely exists. It is not
difficult to derive \cite{BE} that $\gamma = \delta^{}_{0} -
\frac{1}{2} (\delta^{}_{1} + \delta^{}_{-1})$, so that the
corresponding diffraction measure is 
\[
   \widehat{\gamma^{}_{w}} \,=\,
    \bigl( 1 - \cos(2 \pi k) \bigr) \lambda\, .
\]
This is another example of a purely absolutely continuous diffraction
measure. The Radon-Nikodym density relative to $\lambda$ is written as
a function of $k$. In contrast, the dynamical spectrum of this system
contains eigenvalues, wherefore this is an analogue of the Thue-Morse
system \cite{ME}, this time in the presence of absolutely continuous
spectra.

The difference can be rectified by a block map that is very similar
to the map $\phi$ encountered in \eqref{eq:block-map}. Defining $u_{i}
= - w_{i} w_{i+1}$ for $i\in\mathbb{Z}$ maps $w$ to a new sequence
$u$, which almost surely has the diffraction
\[
     \widehat{\hspace{1pt}\gamma^{}_{u}} \, = \,
     \frac{1}{4} \delta^{}_{\mathbb{Z}/2} + \frac{1}{2} \lambda
\]
of mixed type \cite{BE}. In particular, it
displays the entire point part of the original dynamical spectrum at
once, as does the pure point diffraction of the period doubling system
\cite{BMS,BG11} for the Thue-Morse case \cite{BG10}.

\subsection{Renewal processes}

A large and interesting class of processes in one dimension can be
described as a renewal process \cite{BBM,BK}.  Here, one starts from a
probability measure $\mu$ on $\mathbb{R}_{+}$ (the positive real line)
and considers a machine that moves at constant speed along the real
line and drops a point on the line with a waiting time that is
distributed according to $\mu$.  Whenever this happens, the internal
clock is reset and the process resumes. Let us (for simplicity) assume
that both the velocity of the machine and the expectation value of
$\mu$ are $1$, so that we end up with realisations that are, almost
surely, point sets in $\mathbb{R}$ of density $1$ (after we let the
starting point of the machine move to $-\infty$, say).
 
Clearly, the resulting process is stationary and can thus be analysed
by considering all realisations which contain the origin.  Moreover,
there is a clear (distributional) symmetry around the origin, so that
we can determine the corresponding autocorrelation $\gamma$ of almost
all realisations from studying what happens to the right of
$0$. Indeed, if we want to know the frequency per unit length of the
occurrence of two points at distance $x$ (or the corresponding
density), we need to sum the contributions that $x$ is the first point
after $0$, the second point, the third, and so on. In other words, we
almost surely obtain the autocorrelation
\begin{equation} \label{eq:auto-1}
    \gamma \; = \; \delta_{0} + \nu + \widetilde{\nu}
\end{equation}
with $\nu = \mu + \mu * \mu + \mu * \mu * \mu + \ldots$, where the
proper convergence of the sum of iterated convolutions follows from
\cite[Lemma~4]{BBM}. Note that the point measure at $0$ simply
reflects that the almost sure density of the resulting point set is
$1$. Indeed, $\nu$ is a translation bounded positive measure, and
satisfies the renewal relations (see \cite[Ch.~XI.9]{F2} or
\cite[Prop.~1]{BBM} for a proof)
\begin{equation}\label{eq:ren-rel}
   \nu \; = \; \mu + \mu * \nu \qquad\text{and}\qquad
   (1-\widehat{\mu}\, )\, \widehat{\nu}
    \; = \; \widehat{\mu}\, ,
\end{equation}
where $\widehat{\mu}$ is a uniformly continuous and bounded function
on $\mathbb{R}$. The second equation emerges from the first by Fourier
transform, but has been rearranged to highlight the relevance of the
set $S=\{k \mid \widehat{\mu} (k) = 1 \}$ of singularities.  In this
setting, the measure $\gamma$ of \eqref{eq:auto-1} is both positive
and positive definite.

Based on the structure of the support of the underlying probability
measure $\mu$, one can determine the diffraction of the renewal
process as follows.  Let $\mu$ be a probability measure on
$\mathbb{R}_{+}$ with mean $1$, and assume that a moment of $\,\mu$ of
order $1+\varepsilon$ exists for some $\,\varepsilon > 0$ (we refer to
\cite{BBM} for details on this condition). Then, the point sets
obtained from the stationary renewal process based on $\mu$ almost
surely have a diffraction measure of the form
\[
    \widehat{\gamma} \; = \; 
    \widehat{\gamma}^{}_{\mathsf{pp}}  + (1-h)\,\lambda ,
\]
where $h$ is a locally integrable function on $\mathbb{R}$ that is
continuous except for at most countably many points (namely those
of the set $S$). It is given by
\[
    h(k) \; = \; \frac{2\,\bigl(\lvert\widehat{\mu}
    (k)\rvert^2 - \mathrm{Re} (\widehat{\mu}(k))\bigr)}
    {\lvert 1 - \widehat{\mu} (k)\rvert^2}\, .
\]
see also \cite{BBM} for further results on the basis of this
type of argument. Moreover, the pure point part reads
\[
   \widehat{\gamma}^{}_{\sf pp} \; =
   \begin{cases} \delta^{}_{0} , & \text{if\/ $\mathrm{supp}
     (\mu)$ is not a subset of a lattice}, \\
       \delta^{}_{\mathbb{Z}/b} , & \text{otherwise},
     \end{cases}
\]
where $b\mathbb{Z}$ is the
       coarsest lattice that contains $\mathrm{supp} (\mu)$. 
Proofs can be found in \cite{BBM,BK}.

The renewal process is a versatile method to produce interesting point
sets on the line. These include random tilings with finitely many
intervals (which are Delone sets) as well as the homogeneous Poisson
process on the line (where $\mu$ is the exponential distribution with
mean $1$); see \cite[Sec.~3]{BBM} for explicit examples and
applications.  In particular, if one employs a suitably normalised
version of the Gamma distribution, one can formulate a one-parameter
family of renewal processes that continuously interpolates between the
Poisson process (total positional randomness) and the lattice
$\mathbb{Z}$ (perfect periodic order).

\subsection{Point sets from random matrices}

Another interesting class of random point sets derives from the
(scaled) eigenvalue distribution of certain random matrix ensembles;
see \cite{BK} and references therein. The global eigenvalue
distribution of random orthogonal, unitary or symplectic matrix
ensembles is known to asymptotically follow the classic semi-circle
law. More precisely, this law describes the eigenvalue distribution of
the underlying ensembles of symmetric, Hermitian and symplectic
matrices with Gaussian distributed entries.  The corresponding random
matrix ensembles are called GOE, GUE and GSE, with attached
$\beta$-parameters $1$, $2$ and $4$, respectively. They permit an
interpretation as a Coulomb gas, where $\beta$ is the power in the
central potential; see \cite{AGZ,Mehta} for general background and
\cite{D,F} for the results that are relevant here.

{}For matrices of dimension $N$, the semi-circle has radius
$\sqrt{2N/\pi}$ and area $N$. Note that, in comparison with
\cite{Mehta}, we have rescaled the density by a factor $1/\sqrt{\pi}$,
so that we really have a semi-circle (and not a semi-ellipse). To
study the local eigenvalue distribution for diffraction, we rescale
the central region (between $\pm 1$, say) by $\sqrt{2N/\pi} $. This
leads, in the limit as $N\to\infty$, to an ensemble of point sets
on the line that can be interpreted as a stationary, ergodic point
process of intensity $1$; for $\beta=2$, see \cite[Ch.~4.2]{AGZ} and
references therein for details.  Since the process is simple (meaning
that, almost surely, no point is occupied twice), almost all
realisations are point sets of density $1$.

It is possible to calculate the autocorrelation of these processes, on
the basis of Dyson's correlation functions \cite{D}.  Though the
latter originally apply to the circular ensembles, they have been
adapted to the other ensembles by Mehta \cite{Mehta}.  For all three
ensembles mentioned above, this leads to an autocorrelation of the
form
\begin{equation}\label{eq:Dyson-auto}
    \gamma \, = \, \delta_{0} + 
    \bigl( 1 - f(\lvert x \rvert ) \bigr) \lambda 
\end{equation}
where $f$ is a locally integrable function that depends on $\beta$;
see \cite{BK} for the explicit formulas.

\begin{figure}[t]
\centerline{\includegraphics[width=\textwidth]{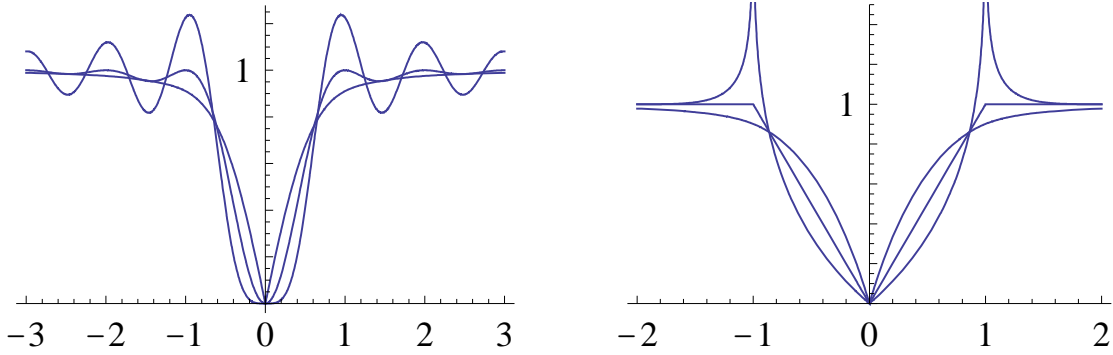}}
\caption{Absolutely continuous part of the autocorrelation (left) and
  the diffraction (right) for the three random matrix derived
  point set ensembles on the
  line, with $\beta\in\{1,2,4\}$. On the left, the oscillatory
  behaviour increases with $\beta$. On the right, $\beta=2$
  corresponds to the piecewise linear function with bends at $0$ and
  $\pm 1$, while $\beta=4$ shows a locally integrable singularity at
  $\pm 1$. The latter reflects the slowly decaying oscillations on
  the left.}  
\label{fig:dyson}
\end{figure}

The diffraction measure is the Fourier transform of $\gamma$, which
has also been calculated in \cite{D,Mehta}. Observing
$\widehat{\delta_{0}} =\lambda$ and $\widehat{\lambda} = \delta_{0}$,
  the result is always of the form
\begin{equation}\label{eq:Dyson-Fourier}
   \widehat{\gamma}\, = \,\delta_{0} + 
   \bigl( 1 - b(k) \bigr) \lambda \,=\,
   \delta_{0} + h(k) \, \lambda ,
\end{equation}
where $b = \widehat{f}$. The Radon-Nikodym density $h$ depends on
$\beta$ and is summarised in \cite{BK}.  Figure~\ref{fig:dyson}
illustrates the result for the three ensembles. 

A similar approach is possible on the basis of the eigenvalues of
general complex random matrices. This leads to the ensemble studied by
Ginibre \cite{Mehta}, which is also discussed in \cite{BK}. One common
feature of the resulting point sets is the effectively repulsive
behaviour of the points, which leads to the `dip' around $0$ for
$\widehat{\gamma}$.

\subsection{Random clusters}

Let us continue by considering the influence of randomness on the
diffraction of point sets and certain structures derived from them in
Euclidean spaces of arbitrary dimension. Here, we start from a single
point set $\varLambda\subset\mathbb{R}^d$, which is then randomly
modified by replacing each point by a (possibly complex) finite random
cluster. This situation is still manageable (via the SLLN) when
$\varLambda$ is sufficiently `nice', for instance if $\varLambda$ is
of finite local complexity and possesses an autocorrelation, which is
then of the form $\gamma= \sum_{z\in\varLambda - \varLambda} \eta (z)
\, \delta_{z}$. More generally, one can analyse this situation in the
setting of stationary ergodic point processes
\cite{G04,BBM,LM,DM08,L09}, which treats almost all realisations at
once and permits a larger generality for the sets $\varLambda$, though
the clusters will then be restricted to positive or signed measures.

Given such a point set $\varLambda$, its (deterministic) Dirac comb
$\delta^{}_{\!\varLambda}$ is turned into a random Dirac comb
\[
   \delta^{(\varOmega)}_{\varLambda} \, = \, \sum_{x\in\varLambda}
    \varOmega_{x} * \delta_{x}
\]
by means of the i.i.d.\ family of random measure
$(\varOmega_{x})_{x\in\varLambda}$ with common law $Q$ and
representing random variable $\varOmega$. Here, we assume that the
expectation $\mathbb{E}_{Q} \bigl( \lvert \varOmega\rvert\bigr)$ is a
finite measure, and that $\mathbb{E}_{Q} \bigl( (\lvert
\varOmega\rvert (\mathbb{R}^d))^{2}\bigr) < \infty$.  Under some mild
(but somewhat technical) conditions \cite{BBM}, one now obtains the
autocorrelation
\[
  \gamma^{(\varOmega)} 
  \, \stackrel{\text{(a.s.)}}{=} \,  
   \bigl( \mathbb{E}_{Q} (\varOmega) * 
   \widetilde{\mathbb{E}_{Q} (\varOmega)} \bigr) * \gamma
   \, +\, \mathrm{dens} (\varLambda)\, \bigl(
   \mathbb{E}_{Q} (\varOmega * \widetilde{\varOmega}) -
   \mathbb{E}_{Q} (\varOmega) * 
   \widetilde{\mathbb{E}_{Q} (\varOmega)} \bigr) * \delta_{0}
\]
and hence the diffraction
\[ 
  \widehat{\gamma}^{(\varOmega)} 
  \, \stackrel{\text{(a.s.)}}{=} \,
   \big\lvert \mathbb{E}_{Q} (\widehat{\varOmega}) 
   \big\rvert^{2}\! \cdot \widehat{\gamma}
   \, + \,  \mathrm{dens} (\varLambda)\, \bigl(
    \mathbb{E}_{Q} ( \lvert \widehat{\varOmega}\rvert^{2}) -
    \lvert \mathbb{E}_{Q} (\widehat{\varOmega}) \rvert^{2}
    \bigr)  \cdot \lambda\, .
\]
The diffraction of the modified structure emerges from the original
one by a modulation of $\widehat{\gamma}$ and the addition of an
absolutely continuous contribution, which in essence is the Fourier
transform of the covariance of the representing random cluster
$\varOmega$.

This approach comprises a wide range of models, including the random
weight and the random displacement model \cite{Hof2} as well as
decorations by random clusters. As mentioned above, the next level of
generality replaces the deterministic set $\varLambda$ by a general
ergodic point process \cite{G04}. This way, both the underlying set
(core process) and the modification (cluster process) are described in
terms of stochastic processes; see \cite{DVJ1,DVJ2} for general
background and \cite{BBM} and references therein for details. This
point of view is a promising starting point for further
investigations.

\section{Concluding remarks}

Our informal exposition was meant to demonstrate that mathematical
diffraction theory provides useful tools for the analysis of
deterministic and random systems, both for practical applications in
crystallography and materials science, and for theoretical questions in
harmonic analysis and dynamical systems theory. While the majority of
the crystallographic literature concentrates on the pure point case,
we have shown that also continuous spectra are explicitly accessible
in relevant cases, and certainly deserve more attention from this
point of view. Merging methods from harmonic analysis and dynamical
systems with well-established procedures from point process theory
might be a good path to proceed. However, this does not only concern
structures with some degree of disorder; there are also classes of
completely deterministic systems whose diffraction is still not fully
understood. A prominent example is provided by tilings with continuous
symmetries such as the pinwheel tiling \cite{R94}, which has
circularly symmetric diffraction \cite{R97,MPS06} that resembles a
powder diffraction pattern \cite{BFG07a,BFG07b}.

The inverse problem of structure determination is already a formidable
problem in the realm of pure point diffractive systems, due to the
existence of non-trivially homometric structures.  As we have shown
above, the inverse problem becomes significantly more involved in the
presence of disorder, including the potential insensitivity to
quantities such as entropy. Although there is quite some knowledge
about this problem in the point process community, it is fair to say
that solutions to the inverse problem or satisfactory classifications
of homometry classes are not in sight.

The setting is by no means restricted to lattice systems, which were
mainly chosen for ease of presentation and concreteness of results.
Also extensions to higher dimensions are possible, where one has to
expect new phenomena (such as the lower rank entropy) that further
complicate the picture. Although the general theory of point processes
is highly developed, the treatment of stochastic systems with
interaction becomes difficult as soon as concrete results are desired.
This is already the case for general equilibrium systems, as they
require the full machinery of Gibbs measures
\cite{Kuel03a,Kuel03b,Kai,BZ08,Z}. Their analysis from the point of
view of mathematical diffraction theory is still in its infancy,
although examples such as the Ising lattice gas show (see \cite{BH}
and references therein) that explicit results are possible.

\end{document}